\documentclass[prd, longbibliography, nofootinbib,floats, superscriptaddress,eqsecnum, tightenlines, showpacs,12pt]{revtex4-1}

\usepackage{amsfonts,amssymb,amstext,amsmath, amsthm,bm,slashed}
\usepackage[dvips]{graphicx}
\usepackage{mathtools}
\usepackage{color}
\usepackage[hidelinks]{hyperref}
\usepackage{tabularx}
\usepackage{mathrsfs}
\theoremstyle{definition}

\theoremstyle{plain}

\usepackage{cleveref}
\usepackage{relsize}

\newcommand{\be}{\begin{equation}}
\newcommand{\ee}{\end{equation}}
\newcommand{\barray}{\begin{array}}
\newcommand{\earray}{\end{array}}
\newcommand{\bea}{\begin{eqnarray}}
\newcommand{\eea}{\end{eqnarray}}
\newcommand{\bs}{\begin{subequations}}
\newcommand{\es}{\end{subequations}}
\newcommand{\beal}{\begin{align}}
\newcommand{\eeal}{\end{align}}

\newcommand{\Ref}[1]{(\ref{#1})}

\def \lb {\left(}

\def\bs{\bar{\sigma}}

\def\pa{\partial}

\newcommand{\rd}{\mathrm{d}}
\newcommand{\p}{\partial}

\newcommand*\overbar[1]{%
  \hbox{%
    \vbox{%
      \hrule height 0.5pt 
      \kern0.4ex
      \hbox{%
        \kern 0em
        \ensuremath{#1}%
        \kern 0em
      }%
    }%
  }%
}

\usepackage{bbm}

\def\g{\mathfrak{g}}

\newcommand{\nnn}{\nonumber\\}

\def\g{\sqrt{|g|}}
\def\l{{{\ell}}}
\def\lb{{\bar\ell}}
\def\const{\text{const.}}
\def\rd{{\mathrm{d}}}

\def\bb{{\bar\beta}}

\begin{document}

\title{Gravity Degrees of Freedom on a Null Surface}

\author{Florian Hopfm\"uller}
\email{fhopfmueller@perimeterinstitute.ca}
\affiliation{Perimeter Institute for Theoretical Physics\\
31 Caroline St N, Waterloo, ON N2L 2Y5, Canada}

\author{Laurent Freidel}
\email{lfreidel@perimeterinstitute.ca}
\affiliation{Perimeter Institute for Theoretical Physics\\
31 Caroline St N, Waterloo, ON N2L 2Y5, Canada}

\date{\today}

\begin{abstract}
A canonical analysis for general relativity is performed on a null surface without fixing the diffeomorphism gauge, and the canonical pairs of configuration and momentum variables are derived. Next to the well--known spin--2 pair, also spin--1 and spin--0 pairs are identified.  The boundary action for a null boundary segment of spacetime is obtained, including terms on codimension two corners.
\end{abstract}

\maketitle

\newpage
\thispagestyle{empty}
\tableofcontents
\newpage

\section{Introduction}

It has long been recognized that the canonical structure of gravity is especially simple on null surfaces. Sachs \cite{sachs62:characteristic_initial_value_problem_gr} was the first to realize that the initial data could be given in an unconstrained form using a double null sheet as an initial value surface. Ashtekar et al. studied the canonical quantization of radiative modes at null infinity \cite{Ashtekar:1981bq}. Epp in \cite{Epp:1995uc} was the first one to write down a proposal for  the null canonical pairs of configuration and momentum variables in the 2+2 formalism (see \cite{brady96:twoplustwo}). 
An additional investigation using stretched horizon techniques was done by Parikh and Wilczek in \cite{Parikh:1997ma}, and investigations in  special coordinates adapted to the null initial data focusing on the construction of the symplectic potential were done by Reisenberger in \cite{reisenberger12:sympl_form_free_null_initial_data}.
More recently Parattu et al. \cite{parattu16:unified_bdy_term, parattu16:bdy_term_grav_action_null} have reconsidered this analysis, focusing on the construction of the analog of the Gibbons--Hawking boundary term for null boundaries, while Lehner et al. in \cite{Lehner:2016vdi} developed further the null boundary action formalism and included null corner terms that generalize Hayward's construction \cite{Hayward:1993my} (see also \cite{Hawking:1996ww} and some preliminary work by Neiman on null corner terms \cite{Neiman:2013lxa}).
  
Our work improves on these latest developments and gives an independent derivation of some of the previous results. Our derivation identifies clearly what the boundary action and the null symplectic structure are without having to recourse to a choice of gauge fixing. In contrast, most approaches gauge fix parts of the diffeomorphism gauge at the beginning of the analysis. This is problematic in our setting for two reasons. Firstly, one has to make an arbitrary choice which clutters the covariance of the final expression. Secondly, we now understand that in the presence of boundaries we cannot fix diffeomorphisms without risking killing key boundary degrees of freedom as explained in \cite{donnelly16:local_subsystems}. Since we want to use  our work in the future to shed some light on the issue of soft gravity modes which we expect to be related to such boundary degrees of freedom, we have to be careful in not making any assumption that will negate the existence of such degrees of freedom. The final decision of whether a mode is physical {\it cannot} be decided beforehand, it is entirely decided by whether it enters non trivially or not in the symplectic structure.

Let us emphasize that understanding the nature of the symplectic structure on a null surface is important for three separate reasons. First, since the initial data is constraint free on null surfaces it can play a fundamental role in understanding the nature of quantum gravity degrees of freedom. Understanding what are the canonical pairs is a necessary step in this direction. 
Moreover, we can also understand the null symplectic potential as a term controlling the flow of information across a null surface \cite{Bousso:1999xy} and as such it is a key element in understanding what could be an appropriate definition of informational horizons and in proving a generalization of the second law associated with finite regions.

Finally,  the understanding of the gravitational null symplectic potential goes hand in hand with the construction of  the boundary action and the corresponding corner terms. These action terms enter the Hamilton--Jacobi function and are the classical analog of the quantum gravity S--matrix.  They therefore play a key role in the interpretation of the physics of gravity inside a finite region when this region possesses null boundaries (see \cite{Brown:2015lvg, Lashkari:2016idm,Hawking:2016msc}).
    
It is thus of great interest to obtain the null canonical structure of gravity without introducing any gauge fixing. We accomplish this here, using a robust, physically intuitive framework, by evaluating the symplectic potential $\Theta_B$ and the corresponding boundary action $A_B$ on a null surface $B$, restricted to variations that preserve the nullness of $B$. 

Let us state our results. We recall that the symplectic potential is schematically of the form $\Theta_B = \int_B P \delta Q+ \int_{\pa B} p \delta q + \delta A_B +\delta a_{\pa B} $ where $(Q,q)$ are the bulk and boundary configuration variables, $(P,p)$ are the corresponding momenta, $A_B$ is the boundary action and $a_{\pa B}$ is the corner action. We assume that the null hypersurface $B$  possesses a ruling of equal time slices $\phi_0=\const$ which define the $(D-2)$ dimensional spacelike cross--sections $S$ of $B$ ($D$ is the space-time dimension). 
 The null geometry of $B$ is encoded into the pair $(q_{ab}, L^a)$, where $L=L^a\pa_a$ is a null vector tangential to $B$ such that $L(\phi^0) =1$, and $q_{ab}$  is the  metric  induced on  the surfaces $S$. It is convenient to decompose the metric $q$ into the product of a  conformal factor $\sqrt{q}^{\ 2/(D-2)}$ and the conformal metric $\tilde{q}_{ab}$ with unit determinant.

One of the non--trivial features of this construction is that it is possible (see also \cite{Lehner:2016vdi}) to choose a boundary action such that the configuration variables $(Q,q)$ depend {\it only} on the null geometry $(q_{ab},L^a)$.
In units where $8\pi G=1$, we find that the null symplectic potential $\int_B P \delta Q$  is
\begin{align}\label{eq:pairs_intro}
\Theta_B^{\text{bulk}} ={}& 
\int_B   \left(\frac12 \tilde\sigma^{ab} \delta \tilde q_{ab} - \bar\eta_a \delta L^a - \left(\kappa+ \tfrac{D-3}{D-2} \theta\right) \delta \ln \sqrt q \right) \rd B.
\end{align}
where $\rd B = \rd \phi^0\wedge \rd S$ is the volume element.
The configuration variables $(\tilde q_{ab},  L^a,\sqrt q)$ describe the induced geometry of the surface $B$, and their conjugate momenta are given by $(\frac12\tilde{\sigma}^{ab}, -\bar{\eta}_a,\left(\kappa+ \tfrac{D-3}{D-2} \theta\right))$. That is our central result.

We thus recover the well--known fact that the momentum of the conformal induced metric $\tilde q$ of a cross--section $S$ of $B$ is the conformal shear $\tilde \sigma^{ab}=-\frac12 q^a{}_{a'} q^b{}_{b'}( {\cal L}_L \tilde{q}^{a'b'})$, a fact first established by Ashtekar et al. \cite{Ashtekar:1981bq} in the context of asymptotic null infinity. 
However, we also see that this symplectic structure involves spin $1$ (i.e., $\delta L^a)$  and spin $0$ modes (i.e.,  $\delta \sqrt{q})$, which are usually and unfortunately gauge fixed away in most treatments. These are exactly what are usually\footnote{This denomination is very confusing since it suggests that soft graviton modes are spin two degrees of freedom, while they are in fact a combination of spin $1$ and spin $0$ modes. This fact, which is usually misunderstood, will be expanded on in \cite{hopfmueller:soft_gravitons}.} called soft graviton modes. What our results show is that since the spin $0$ and $1$ modes enter the symplectic structure they are physical degrees of freedom. They cannot be gauge fixed away, but only pushed to the corners of $B$ by applying a diffeomorphism. 

The momentum of the normal vector field $L$ to $B$ is the \emph{twist} $\bar\eta$ defined as 
\begin{equation}
\bar\eta_a = - q_a{}^b \nabla_L \bar L_b.
\end{equation}
Here, $q_a{}^b$ is the projector onto $S$, and $\bar L$ is a null form that is orthogonal to $S$ and normalized as $L^a \bar L_a = 1$. The twist is thus the parallel transport of the auxiliary null form $\bar{L}$ along $L$, and describes how the surfaces $S$ twist inside $B$ when one follows the integral curves of $L$. It is closely related to Damour's momentum \cite{damour82:surface} that appears in the study of stretched horizons.
Lastly, the momentum of the volume element $\sqrt q$ of a cross--section of $B$ is a combination of the \emph{expansion} $\theta=q^{ab}\nabla_aL_b$ along $L$ and the \emph{surface gravity} $\kappa$ which enters in $\nabla_L L^a=\kappa L^a$. 

The form of the symplectic potential given above depends on what the precise action for gravity is when space--time has a null boundary. When the cosmological constant vanishes, the on--shell action is a pure boundary integral and its value is the Hamilton--Jacobi functional and thus of great importance. We give a null boundary action that includes corners and generalizes  the result of \cite{Lehner:2016vdi}. 
One of the key elements entering the boundary action is the surface gravity $\kappa$ already encountered, while the key element entering the corner term is the factor $h$, which is the (logarithmic) normal volume element: $e^h= \sqrt{|g|}/\sqrt{q}$. It thus measures the size of the normal geometry, and can be physically identified as the {\it redshift factor}. As we will see in section \ref{redshift}, it is proportional to the redshift experienced by light rays skimming along $B$ as measured by geodesic observers crossing $B$. Our result for the boundary plus corner action is 
\begin{equation}
A_{B} + a_{\pa B}  = \int_B \kappa \rd B + \frac12 \int_{\p B} (1-h) L^a \rd_a S,
\end{equation}
where $\rd_a S:=i_{\pa_a} \rd B$ is the directed volume element on $\p B$. 

Finally, we also determine the corner symplectic structure $\Theta_{\p B} = \int_{\p B} p \delta q$.
Its general expression is given later (\ref{eq:Theta_terms}), but if we assume that the boundary of $B$ consists of an initial and a final sphere at constant $\phi^0$:  $\pa B = S_1\cup S_0$, then it simply reads:
\be 
\Theta_{\partial B} ={}  \frac12  \int_{S_0}^{S^1}  \big(1+h\big)  \delta \ln\sqrt{q}  \, \rd S 
\ee 
This shows that the redshift factor $h$ is a variable conjugate to the angular size $\sqrt{q}$.
This is an interesting relationship especially in view of the 
Etherington reciprocity law relating the area distances to the redshift factor \cite{Etherington}.

The remainder of this article is organized as follows. In section \ref{sec:symp_geo} we briefly review the symplectic geometry of field space in order to fix our definitions and notation. Section \ref{sec:setup} contains our geometrical setup. In section \ref{sec:Theta_null} we perform our central calculation, obtaining the null canonical pairs of gravity in section \ref{sec:canonical_pairs}. Section \ref{sec:lagrangian_bdy} contains our suggestion for a Lagrangian boundary term. We conclude in section \ref{sec:conclusion}, and collect some of the more technical calculations in the appendices.

\section{The Symplectic Geometry of Field Space}\label{sec:symp_geo}


The pre--symplectic geometry of field space can be obtained in a covariant way (see, e.g., \cite{crnkovic87, leewald, donnelly16:local_subsystems}), which we briefly review here. It is described by the pre--symplectic form $\Omega_B$, which is a closed two--form on field space and an integral over a Cauchy hypersurface $B$ in space--time.  The prefix ``pre'' refers to the fact that $\Omega_B$ on field space has degenerate directions, so it does not qualify as ``symplectic''. The degenerate directions are the gauge degrees of freedom, which  have to be ultimately quotiented out to obtain the physical phase space. When the background fields are taken to be on--shell,  $\Omega_B$ is independent of the choice of $B$ and only depends on its homology class. Here, we specialize to the case that the hypersurface $B$ is null.

Schematically, $\Omega_B$ can be written as
\begin{equation}
 \Omega_B = \int_B \delta P \curlywedge \delta Q + \int_{\partial B} \delta p \curlywedge \delta q.
\end{equation}
Here, $\delta$ is the exterior derivative on field space, and $\curlywedge$ is the wedge product on field space. The pairs $(Q, P)$ of configuration and momentum variables are the canonical pairs. We have allowed for the presence of \emph{corner} degrees of freedom $(q,p)$. Note that in a Hamiltonian analysis, ``corner'' refers to the boundary of the hypersurface $B$, i.e., it is a codimension two surface. 
In the following and for gravity  we will focus on the case where the configuration variables $(Q,q)$ are linear functionals of the metric while the momenta include derivatives of the metric.

$\Omega_B$ is the field space exterior derivative $\Omega_B = \delta \Theta_B$ of the \emph{symplectic potential} $\Theta_B$. 
The symplectic potential $\Theta_B$ is the integral of the \emph{symplectic potential current} $\Theta$, which is a one--form on field space and a $(D-1)$--form on spacetime. $\Theta$ is obtained through the equation
\begin{equation}\label{eq:def_eq_Theta}
 \delta L =:  \rd \Theta- E.
\end{equation} 
Here $L$ is the Lagrangian density, which is a $D$--form on space--time. $E$ are the equations of motion, which are a one--form on field space and a $D$--form on space--time. By definition they do not contain derivatives of the variations of the fields, and they are uniquely determined by the Lagrangian. Here $\rd$ is the space--time exterior derivative. \Ref{eq:def_eq_Theta} determines $\Theta$ only up to the addition of a closed $(D-1)$--form on $M$. This ambiguity can be fixed by demanding the consistency of the variation for boundaries including corners (see \cite{Freidel:2013jfa}).
Schematically the symplectic potential is of the form
\begin{align}\label{eq:Theta_schematic}
 \Theta_B ={} \Theta_B^{\text{bulk}} +  \Theta_{\partial B}  + \delta A_B + \delta a_{\partial B},
\end{align}
where 
$
\Theta_B^{\text{bulk}} = \int_B P \delta Q$ and $ \Theta_{\p B}= \int_{\partial B} p
\delta q.$
The total variation terms $\delta A_B$ and $\delta a_{\partial B}$ do not contribute to the symplectic form, because $\delta \delta = 0$. These terms can be reabsorbed into a redefinition of the action $S \to S - A_B -a_{\pa B}$. The inclusion of these terms corresponds to a choice of polarization, and is necessary if one demands that the configuration variables do not include metric derivatives.
Finally an important point is that 
we assumed the field space exterior derivative $\delta$ and the integral $\int_B$ commute. That means the location of the hypersurface $B$ must be specified in a field independent way.

Let us specialize to our case of vacuum metric general relativity without cosmological constant. The Lagrangian density $L$, the equations of motion $E$ and the symplectic potential current $\Theta$ are:
\begin{align} \label{eq:Theta_general}
L = \frac12 R \epsilon, \qquad E = \frac12 G^{ab} \delta g_{ab} \epsilon, \qquad  \Theta = \frac12 \nabla_b ( \delta g^{ab} - g^{ab} \delta g) \epsilon_a.
\end{align}
$\delta g^{ab} = g^{ac} g^{bd} \delta g_{bd}$ denotes the metric variation with indices raised, not the variation of the inverse metric. $\delta g = g^{ab} \delta g_{ab}$ is its trace. We set $8 \pi G = c = 1$, and introduced the volume $D$ form and the directed volume $(D-1)$--form
\begin{equation} \label{eq:volume_forms}
 \epsilon = *1 = \frac{\g}{D!} \epsilon_{a_1 ... a_D} \rd x^{a_1} \wedge ... \wedge \rd x^{a_D}, \qquad \epsilon_a = \iota_{a} \epsilon =  \frac{\g}{(D-1)!} \epsilon_{a a_2 ... a_D} \rd x^{a_2} \wedge...\wedge\rd x^{a_D}.
\end{equation}
We made the usual choice in fixing the space--time closed ambiguity in $\Theta$ and used the standard expression (see, e.g., \cite{Iyer1994}). The analysis of the closed ambiguity will be part of future work.

We  see from \Ref{eq:Theta_general} that the covariant symplectic potential contains variations of metric derivatives.
The challenge we face is to express it  purely in terms of variations of the metric only, so that we can read off the proper canonical pairs of gravity. We therefore have to manipulate $\Theta_B$ to remove the derivatives of variations.
The derivatives of variations will be of two kinds: Derivatives tangential to $B$, and derivatives in directions transverse to $B$. 
The tangential derivatives can easily be taken care of by integrating by parts. The transverse derivatives are more subtle, but we will show that they can be eliminated through variation by parts, i.e., they can be absorbed into a total variation. Carefully carrying through this procedure and keeping all the terms 
is the first goal of this paper. It will give us an expression for $\Theta_B$ as the sum of a bulk term, a boundary term, a bulk total variation and a boundary total variation.

\section{Setup} \label{sec:setup}
In this section, we introduce the structures and notation we will use to evaluate the symplectic potential on the null hypersurface $B$. The setup is taken from \cite{freidel16:twoplustwo} and \cite{hopfmueller16:null_bdies_cov_ham_gr}. Previous, similar formalisms were set up e.g. in \cite{Hayward:1993dq, brady96:twoplustwo, Gourgoulhon:2005ng}. 

\subsection{Foliations, Normal Forms and Coordinates}

\begin{figure}[t]
\centering
\includegraphics{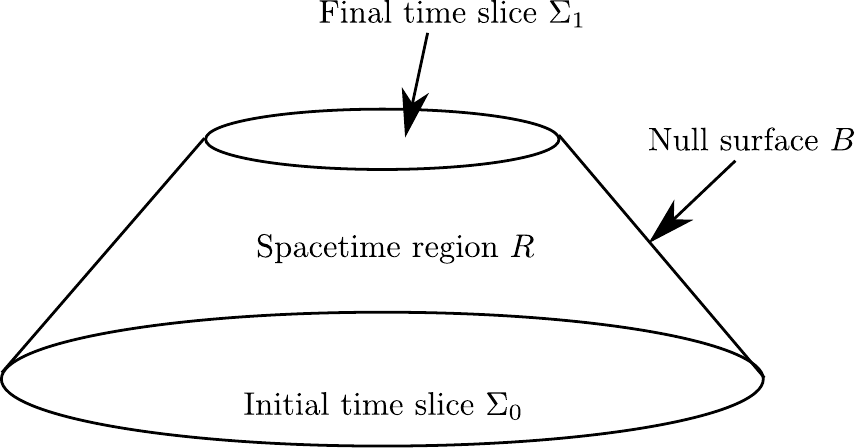}
\caption{A typical situation where the symplectic structure on the null surface $B$ is of interest is when $B$ is part of the boundary of the spacetime region $R$ under consideration. The other parts of the boundary are spacelike surfaces $\Sigma_i$.}
\label{fig:typical_spacetime}
\end{figure}

Let $M$ be the $D$--dimensional space--time.
We are typically interested in a region $R$ of a $D$- dimensional space--time with boundary $B\cup \Sigma_0\cup\Sigma_1$ where $\Sigma_i$ are spacelike hypersurfaces and $B$ a null hypersurface (see figure \ref{fig:typical_spacetime}).
More generally we want to understand the nature of the  symplectic potential $\Theta_B$ on a null hypersurface $B$. The location of $B$ is specified by the condition $\phi^1(x) = 0$, where $\phi^1$ is a suitable scalar field on $M$ that increases towards the past of $B$. $B$ is a finite hypersurface with a boundary $\partial B$ that we will call a ``corner''. It is a member of the foliation specified by $\phi^1 = \const$ and located at $\phi_1=0$. We do not assume that every member of the foliation is a null hypersurface, but assume that $\phi^1$ is a good foliation function in a neighbourhood of $B$, i.e., $\rd \phi^1 \neq 0$ on $B$.

We introduce another foliation given by $\phi^0 = \const$ of spacelike  hypersurfaces, where $\phi^0$ is a field that increases towards the future. We require that $\phi^0$ is a good foliation function in a neighbourhood of $B$, and that nowhere $\rd \phi^0$ is a multiple of $\rd \phi^1$. 
At the intersections of the two foliations lies a two--parameter family of spacelike
codimension $2$ surfaces $S$. 
Coordinates $\sigma^A(x)$ are also chosen on each surface $S$. They are not required to be constant on the null generators of $B$. Doing so would be a partial gauge fixing which we want to avoid, since the direction of the null generators is metric dependent. 
 
Using also the foliation fields as coordinates, we introduce a frame $(X^a)(x) = (\phi^i, \sigma^A)(x)$ on $M$.
This frame represents an invertible mapping $X: U \to M$, from a domain $U\in \mathbb{R}^D$ to $M$. 
The metric $G$ on $M$ can be represented as a metric on $U$ via the pullback: 
$X^*G =g$.
Here and in the following, $a \in \{0, ..., D-1\}, i \in \{0,1\}$ and $A \in \{2, ...,  D-1\}$. $x$ represents a choice of coordinates while $X^a(x)$ represents points of $M$. We will refer to $X^a$ as a \emph{foliation frame}. We introduce the separation of foliation frame and coordinates also with an eye on future work in order to have full control of what we vary and what not. The setup also connects to the formalism developed in \cite{donnelly16:local_subsystems}, where the frame fields become physical.
In the foliation frame, the tangent vectors $e_A$ to the surfaces $S$ become
\begin{equation}
 e_A = e^a_A \partial_a, \text{ where } e^a_A = \frac{\partial x^a}{\partial \sigma^A} = \delta^a_A,
\end{equation}
while the metric  in the foliation frame can be parametrised as
\begin{align}\label{eq:metric_param}
 \rd s^2 = g_{ab} \rd x^a \otimes \rd x^b = H_{ij} \rd \phi^i \otimes \rd \phi^j + q_{AB} (\rd \sigma^A - A^A_i \rd \phi^i) \otimes (\rd \sigma^B - A^B_j \rd \phi^j).
\end{align}
Here we have defined the \emph{shift connection} $A^A:= A_i^A \rd \phi^i $, which is a one--form in the normal plane to $S$ valued into  $TS$. We also defined the \emph{normal metric} $H_{ij}$, which determines the geometry of the normal two--planes $(TS)^\perp$ to $S$,
while $q$ is the {\it tangential metric} which determines the geometry of the sphere $S$. The metric $g$ contains $\frac12 D(D+1)$ parameters and this parametrization is completely general. No gauge fixing has taken place, and we have not yet specialized to the case of a null hypersurface $B$.  Gauge fixing at this stage risks killing physical degrees of freedom.
This important point is, in most previous approaches, completely neglected and leads to a deep source of confusion about what is physical and what is not.

The inverse metric is
\begin{align}
  g^{ab} \partial_a \otimes \partial_b ={} & H^{ij}(\partial_i + A_i^A \partial_A)\otimes(\partial_j + A_j^B \p_B) + q^{AB} \partial_A \otimes \partial_B,
\end{align}  
where $H^{ij}$ and $q^{AB}$ are the inverses of $H_{ij}$ and $q_{AB}$, respectively. 
We introduce the covariant normal derivatives
 \be \label{eq:shifted_derivative}
 D_i :=  (\partial_i + A_i^A \partial_A).
 \ee
They can be understood as normal derivatives covariant under the gauge group $\text{Diff}(S)$ of diffeomorphisms on $S$. That is because under an infinitesimal change of foliation frame $\delta_V \phi^i =0$ and $\delta_V \sigma^A = V^A(x)$, the normal metric transforms as a scalar $\delta H_{ij}=V^C\pa_C H_{ij}$, the tangential metric transforms as a tensor $\delta_V q_{AB} = {\cal L}_V q_{AB}$, while  $A_i^A$ transforms as a connection:
\be \label{eq:normal_connection}
 \delta_V A_i^A = \partial_i V^A +[ A_i, V]_S^A,
\ee
where $[., .]_S$ is the Lie bracket on $S$. Then, the derivative $D_i$ transforms covariantly as a scalar under the gauge group $\text{Diff}(S)$: $\delta_V (D_i f) = V^A \partial_A (D_i f)$ for a field $f$ on $M$.
The curvature of the normal connection is the vector field 
\be
[D_0,D_1]^A= \pa_0 A_1^A - \pa_1 A_0^A + [A_0,A_1]_S^A.
\ee
We introduce the logarithmic normal volume element $h$ as
\begin{equation}
e^h = \sqrt{|H|}.
\end{equation}
It will play a very important role in the symplectic structure and the boundary action and will be interpreted as the \textit{redshift factor} in an adapted frame. 
The determinants of the normal metric $H_{ij}$, the induced metric $q_{AB}$ and the full metric $g_{ab}$ are therefore linked by
\begin{equation}
 \sqrt{|g|} = e^h \sqrt q.
\end{equation}
In order to write the symplectic potential using quantities intrinsic to the surfaces $S$ we need to be able to project along its two normal directions. We therefore have to choose a basis of  one--forms normal to $S$.  
There is a simple choice of basis which is metric independent, and depends only on the choice of foliation. It is given by $(\rd \phi^0,\rd \phi^1)\in (TS)^\perp$.
However, since the surfaces $S$ are  part of a null hypersurface, the most convenient choice is to use a null co--frame $(\bm\ell,\bm{\bar{\ell}})$ consisting of two null forms normal to the family of surfaces $S$, one of which will be normal also to $B$. This is what we do here.

Let $\bm \l = \l_a \rd x^a$ and $\bm \lb = \lb_a \rd x^a$ be two smooth, null one--form fields normal to the surfaces $S$ (here and in the following bold--face letters denote one--forms). Let $\bm \l$ be such that at $B$, $\bm \l$ is normal to $B$, and $g^{-1} (\bm \l) = \l^a \partial_a$ is future pointing. We impose the normalization condition that $g^{-1}(\bm \l,  \bm \lb) = 1$. These conditions uniquely determine $\bm\l$ and $\bm\lb$ in a neighbourhood of $B$, up to a rescaling $(\bm\l \rightarrow e^\epsilon \bm\l, \bm\lb \rightarrow e^{-\epsilon} \bm\lb)$, where $\epsilon$ is an arbitrary function. Our choice of a null dyad diagonalizes the $SO(1,1)$--symmetry of the normal plane to $S$, and the rescaling is the action of a $SO(1,1)$ transformation.
 Since $(\bm \l , \bm \lb)$ and $(\rd\phi^0,\rd\phi^1)$ both form a basis of $(TS)^\perp$ their relationships  can be parametrized in terms of four fields $\alpha, \bar\alpha, \beta$ and $\bb$ which form a set of generalized lapses. We set 
\begin{align}\label{eq:l_form} 
\bm \l ={}& e^{\alpha} ( \rd \phi^1 - \beta  \rd \phi^0), \nonumber \\
\bm \lb ={}&   \frac{e^{\bar\alpha}}{1+\beta \bb}( \rd \phi^0 + \bb \rd \phi^1).
\end{align}
The condition that the slices $\phi^1=\const$ are timelike or null  and that the slices $\phi^0=\const$ are   spacelike is encoded in the inequalities $\beta\geq 0,\bar\beta>0$. The  four functions $(\alpha, \bar\alpha, \beta, \bb)$ determine the inverse normal metric $H$ through the conditions $H^{ij} = g^{-1}(\rd \phi^i, \rd\phi^j)$. We get

\begin{align} \label{eq:h}
 H^{ij} = \frac{e^{-h}}{1+\beta\bb}
 \begin{pmatrix}
	 -2 \bb & 1 - \beta\bb\\
	 1-\beta \bb & 2 \beta
 \end{pmatrix}, \qquad
 H_{ij} = \frac{e^{h}}{1+\beta\bb}
 \begin{pmatrix}
	 -2 \beta & 1-\beta\bb\\
	 1-\beta\bb & 2 \bb
 \end{pmatrix},
\end{align}
where the normal volume element $h$ is
\begin{equation}\label{eq:metric_determinants}
h = \alpha+\bar\alpha.
\end{equation}

The quantity $\alpha - \bar\alpha$ does not enter the metric, and encodes the rescaling freedom in $\bm \l$ and $\bm \lb$ alluded to above. $\alpha - \bar\alpha$ is therefore not physical, it is pure gauge freedom. We will refer to it as the \emph{boost gauge}, because a boost transformation in the normal plane to $S$ will change $\alpha-\bar \alpha$, keeping $h$ and the directions of $(\bm\l, \bm\lb)$ fixed. A boost transformation $(\l, \lb) \rightarrow (e^\epsilon \l, e^{-\epsilon}\lb)$ acts as $(\alpha, \bar\alpha) \rightarrow (\alpha+\epsilon, \bar\alpha-\epsilon)$.

Even though it is pure gauge, we will not fix $\alpha-\bar\alpha$ for now. In the literature different choices are made, and the generality of our boost gauge allows to connect them. For instance, 
\cite{parattu16:bdy_term_grav_action_null} and the BMS literature work in the gauge $\alpha = 0$ while \cite{Lehner:2016vdi} works in the gauge $\bar\alpha = 0$. We will see that it is more convenient for the problem at hand to choose $\bar{\alpha}=0$ such that $\alpha=h$. Note that the boost gauge can be fixed only with reference to the foliation functions $\phi^0, \phi^1$, and a boost gauge fixing thus depends on how we parametrize the family of surfaces $S$.

\begin{figure}[t]
\centering
 \includegraphics[width=.5\textwidth]{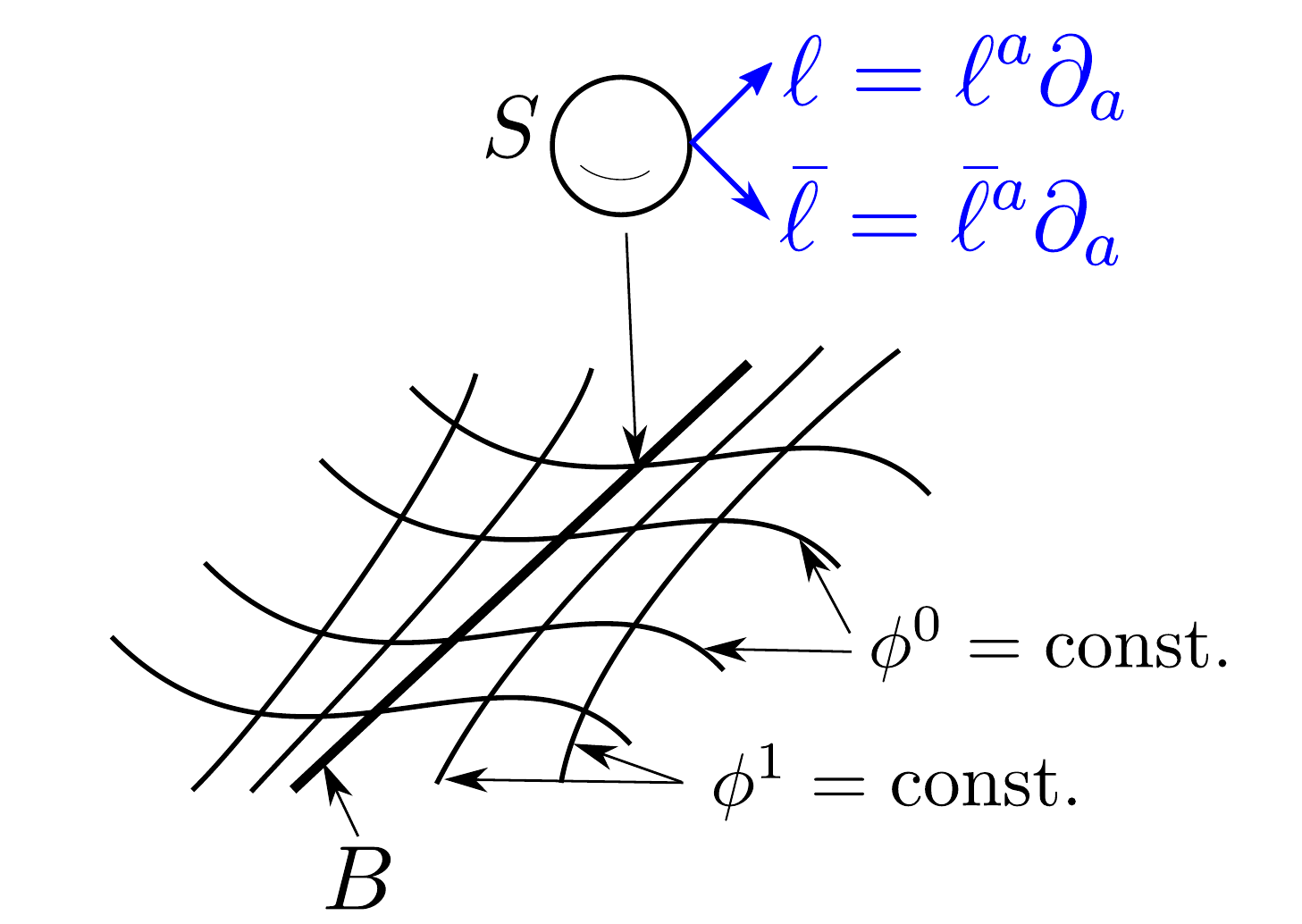}
 \caption{The geometry of our setup is depicted. The null hypersurface $B$ is a member of the foliation $\phi^1 = \const$ that need not be null everywhere. It is ruled into codimension two surfaces $S$ by a second foliation $\phi^0=\const$ The vectors $\l$ and $\lb$ are null and normal to $S$. $\l$ is normal also to $B$, and since $B$ is null it is at the same time tangential to $B$. $\lb$ is transverse to $B$, and the vectors are normalized as $\l^a \lb^b g_{ab} = 1$.}
 \label{fig:geometry}
\end{figure}

While the forms are denoted by bold letters, we denote the corresponding vectors with unbolded letters as $\l = g^{-1} (\bm \l)$ and $\lb = g^{-1} (\bm \lb)$. They are  obtained by raising the index on $\bm \l$ and $\bm \lb$ and are given by
\begin{align}\label{eq:null_vectors}
 \l = \l^a \partial_a = e^{-\bar\alpha} (D_0 + \beta D_1),\qquad \lb =  \lb^a\partial_a = \frac{e^{-\alpha}}{1+\beta\bb}(D_1 - \bb D_0).
\end{align}
Note that the forms $(\bm \l, \bm \lb)$ as well as the vectors $(\l, \lb, D_i)$ contain metric parameters and are thus metric dependent. For an illustration of the geometry, see figure \ref{fig:geometry}.

For notational convenience, we will mostly work with tensors that have $D$--dimensional indices, even if they are intrinsic to $S$. Vectors $v^A$ and contravariant tensors on $S$ are pushed forward into $M$ along the inclusion, yielding in the foliation frame $v^a = e^a_A v^A = \delta^a_A v^A$. Covectors and covariant tensors like $q_{AB}$ are pushed forward using the forms
\begin{equation}
\bm e^A := e^A_a \rd x^a = (q^{AB} g_{ab} e^b_B) \rd x^a = (\delta_a^A - A_a^A) \rd x^a,
\end{equation} 
yielding, e.g., $q_{ab} = (\delta_a^A - A_a^A)(\delta_b^B - A_b^B) q_{AB}$. 

Using that notation, we can write the components of the shifted derivative in foliation coordinates as $D_i^a = \delta_i^a + A^a_i$. It can be checked that $q_{ab} D_i^b = q_{ab} \l^a = q_{ab} \lb^a = 0$. The two vectors $(D_i)$ span the same space as the vectors $(\l, \lb)$, and all four are indeed orthogonal to $S$.
It can easily be checked that the induced metric $q$ on $S$ satisfies the completeness relation 
\begin{equation}\label{eq:completeness}
 q_{ab} + \l_a \lb_b + \lb_a \l_b = g_{ab}.
\end{equation}
$q$ also satisfies $q_a{}^b q_{bc} = q_{ac}$.

So far, the setup we described works for any two foliations $(\phi^0, \phi^1)$ with spacelike intersections. Let us now specialize to the case that $\phi^1 = 0$ describes a null hypersurface $B$. The nullness condition reads $g^{ab} (\nabla_a \phi^1) (\nabla_b \phi^1) =  H^{11} \overset{B}{=} 0$, and from \Ref{eq:h} we see that this is equivalent to the condition $\beta \overset{B}{=} 0$. So we get
\begin{equation}
 \beta \overset{B}{=} 0, \qquad \bm \l \overset{B}{=} e^\alpha \rd \phi^1, \qquad \l \overset{B}{=} e^{-\bar\alpha} D_0.
\end{equation}
Note that also the derivatives $\nabla_\l \beta$ and $q_a{}^b \nabla_b \beta$ vanish on $B$. 
We see that as expected, the vector $\l$ is parallel to $B$ since on $B$ it does not contain a transverse derivative $\partial_{\phi^1}$. Its integral curves are the null generators of $B$. If we had chosen the coordinates $\sigma$ to be constant along the null generators of $B$, then the shift $A_0^a$ would vanish on $B$ yielding $\l^a = e^{\bar\alpha} \p_{\phi^0}$.
The induced metric on $B$ is
\begin{equation}\label{eq:induced_metric}
 \rd s^2 \vert_B = q_{AB} (\rd \sigma^A - A_0^A \rd \phi^0)\otimes(\rd\sigma^B - A_0^B \rd \phi^0),
\end{equation}
where we have used that $h_{00} \overset{B}{=}0$. Its parameters are $q_{AB}$ and $A_0^A$, and the number of parameters is $\frac12 D(D-1) -1$, as expected for the induced metric of a codimension $1$ hypersurface that satisfies one condition. The directed volume element  on $B$ can be written in terms of the volume form $\rd S$ on $S$:  We set $\rd S := \sqrt{q}  \rd^{D-2}\sigma$ and obtain
\begin{equation} \label{eq:volume_form_B}
\epsilon_a \vert_B = \iota_a \epsilon \vert_B 
=- \l_a  e^{\bar\alpha} \rd \phi^0 \wedge \rd S.
\end{equation}
The equality can be seen by writing $\iota_a \epsilon$ from \Ref{eq:volume_forms} in the foliation frame, where $\epsilon = e^h \rd\phi^0\wedge \rd \phi^1 \wedge \rd S$, setting $\rd \phi^1 = 0$, substituting $\iota_a \rd \phi^1 = \partial_a \phi^1 \overset{B}{=} e^{-\alpha} \l_a$, and using $h = \alpha+\bar\alpha$.
The combination $L_a =e^{\bar \alpha} \ell_a$ that enters the induced volume form will play a special role in our construction, as we will see.

The introduction of the auxiliary foliation $\phi^0$ on $S$ should be thought of as a choice of reference frame on $B$. It avoids dealing with the degenerate induced metric on $B$ and makes calculations more straightforward, but comes at the cost of introducing some additional structure into the setup: the decomposition of $B$ into spheres $S$. Note, however, that we need an auxiliary foliation $\phi^0$ in order to locate the  position of the corner $\p B$, so we cannot avoid introducing such extra data, at least near the boundary of $B$. It has been understood recently that when $B$ is a part of the boundary of a space--time region, then the foliation fields $(\phi^0, \phi^1)$ which provide a frame  around $B$ acquire direct physical meaning, as a label of boundary degrees of freedom \cite{donnelly16:local_subsystems}.

\subsection{Decomposition of Metric Variations}
The symplectic potential contains the variation of the space--time metric, $\delta g_{ab}$. For now, we will consider a completely general metric variation, but later we will specialize to the case that the metric variations leave the hypersurface $B$ null. We view the foliations $(\phi^0, \phi^1)$ and the coordinates $\sigma$ as fixed, so they do not vary: $\delta\phi^i=\delta \sigma^A=0$. Since the position of $B$ is described using the foliations, this also ensures that $B$ does not move, while its geometry varies, so that integral signs and variations commute. As before, we write $\delta g = g^{ab} \delta g_{ab}$ for the trace of the metric variation, and $\delta g^{ab} = g^{ac} g^{bd}\delta g_{cd}$ is the variation of the metric with the indices raised. 

The variation of the metric will be decomposed into tensors intrinsic to $S$, using the structure of the two foliations. We then express it using the variations of $q, \l$ and $\lb$. Note that since the forms $(\bm \l, \bm \lb)$ are linear combinations of the $\rd\phi^i$ which do not vary, their variations stays orthogonal to the surfaces $S$, i.e., $q^{ab} \delta \l_b = 0$ (and similarly for $\bm \lb$).
The relationships among $\l$ and  $\lb$, which are implemented by the definition of the metric dependent coefficients $(\alpha,\bar\alpha,\beta,\bar{\beta})$, are also preserved under variations: We have $\delta (\l_a \l^a) = 0$, $\delta (\l^a \lb_a) = 0$, etc.

Our first variation quantity,
\begin{align}
 \delta q_{ab} : = q_a{}^c q_b{}^d \delta g_{ab} = \delta q_{AB}e^A_a e^B_b
\end{align} 
is the variation of the induced metric, pushed forward into $M$. Its trace $\delta q : = q^{ab} \delta q_{ab} = q^{AB} \delta q_{AB}$ is related to the change of the area element on $S$ as $\delta \sqrt q = \frac12 \sqrt q \delta q$. Note that $\delta q_{ab} \neq \delta(q_{ab})$, because the latter expression contains the variation of the tensors $e^A_a$.

The vector $\l$ is null and normal to $S$ by definition, but both of these properties are metric dependent. When the metric varies, $\l$ will therefore change to restore the properties. The change in $\l$ parallel to $S$ is $q^a{}_b \delta \l^b$. It can be written as
\begin{align}
q^a{}_b \delta \l^b = - q^{ab} \l^c \delta g_{bc} = e^{-\bar\alpha} (\delta A_0^a + \beta \delta A_1^a) \overset{B}{=} e^{-\bar\alpha}\delta A_0^a,
\end{align}
For the first identity, we have used $\l^a \delta g_{ab} = \delta \l_b - g_{ab} \delta \l^a$, and that the variation $q_a{}^b \delta \l_b = 0$, since $\bm\l$ is fixed to be normal to $S$. For the second identity, we varied the expression (\ref{eq:null_vectors}), and used that $\delta D_i^a = \delta A_i^a$ and that $D_i^a q_{ab} = 0$. Similarly, we get
\begin{equation}\label{dA1}
q^a{}_b \delta \lb^b = - q^{ab} \lb^c \delta g_{bc} =  \frac{e^{-\alpha}}{1+\beta\bb} (\delta A_1^a - \bb \delta A_0^a).
\end{equation}
The change of the normal volume element $e^h$ is given by $\l^a \lb^b \delta g_{ab}$:
\begin{equation}
 \l^a \lb^b \delta g_{ab}  =  \lb^a \delta \l_a + \l^a \delta \lb_a = \delta (\alpha+\bar\alpha) = \delta h.
\end{equation}
The second equality can be checked explicitly using the expressions (\ref{eq:l_form}) and varying them.
Remembering that $\g = \sqrt q e^h$, and noting that $\delta \g = \frac12 \g \delta g$, we get 
\begin{equation} \label{eq:delta_g}
\delta g = g^{ab} \delta g_{ab} = \delta q + 2 (\lb^a \delta \l_a+ \l^a \delta \lb_a).
\end{equation}
The part of the change in $\l$ that lies in the normal plane to $S$ and is not parallel to $\l$ is given by $\l^a \delta \l_a$.
We obtain
\begin{align}
\l^a \delta \l_a = \frac12 \l^a \l^b \delta g_{ab} = - e^{\alpha-\bar\alpha} \delta \beta,
\end{align}
so on $B$, $\l^a \delta \l_a$ encodes how much $B$ changes away from being null. Similarly, we get 
\be
  \lb^a \delta \lb_a = \frac12 \lb^a \lb^b \delta g_{ab} = \frac{e^{\bar\alpha-\alpha}}{(1+\beta\bb)^2} \delta \bb.
\ee 

We have listed all possible projections of $\delta g_{ab}$ with $\l, \lb$ and $q$, and expressed them as variations of $\l, \lb$ and $q$ and as variations of the metric parameters. Using the completeness relation \Ref{eq:completeness} the metric variation can thus be expressed fully in terms of the variations we have given, as
\begin{align}
 \delta g_{ab} ={}& \delta q_{ab} - (\l_a q_{bc} \delta \lb^c + \l_b q_{ac} \delta \lb^c) - (\lb_a q_{bc} \delta \l^c + \lb_b q_{ac} \delta \l^c) + (\l_a \lb_b + \lb_a \l_b) (\lb^c \delta \l_c + \l^c \delta \lb_c) \nonumber\\
 & + 2 \l_a \l_b (\lb^c \delta \lb_c) + 2 \lb_a \lb_b (\l^c \delta \l_c).
\end{align} 
The change of normalization of $\bm\l$ is $\lb^a \delta \l_a$, and the change of the normalization of $\bm \lb$ is $\l^a \delta \lb_a$. They enter the metric variation only through the symmetric combination $\lb^a \delta \l_a + \l^a \delta \lb_a$. This is the variational expression of the fact that the boost gauge of $\bm\l$ and $\bm \lb$ is indeed pure gauge.

\subsection{Extrinsic Geometry}
The momenta conjugate to the metric are the extrinsic geometry of $S$. As it was the case with our variations, all of the extrinsic geometry is expressed in tensors intrinsic to $S$, which we push forward onto $M$. We will not give a complete list here, but just define the ones that will appear in our calculations.

The \emph{extrinsic curvature} associated with $\l$ is 
\begin{align}
 \theta_\l^{ab} :={}& q^{ac} q^{bd} \nabla_c \l_d = \frac12 q^{ac} q^{bd} \pounds_\l q_{cd}
\end{align}
It describes how the induced two--metric changes along the vectors $\l$ and is symmetric by Frobenius' theorem because $\l$ is surface orthogonal to $S$. Its trace $\theta_\l = q_{ab} \theta^{ab}_\l = q^{ab} \nabla_a \l_b$ is the \emph{expansion}. It measures how the area element $\sqrt q$ on $S$ changes along $\l$, corrected for the divergence of the coordinate lines $\sigma = \const$ relative to $\l$ and the normalization of $\l$, and can be written as
\begin{equation}
 \sqrt q e^{\bar\alpha} \theta_\l = \partial_a (\sqrt q D_0^a) - \beta \partial_a (\sqrt q D_1^a) \overset{B}{=} \partial_a (\sqrt q D_0^a).
\end{equation}
(see appendix \ref{ap:extrinsic}). If the shift $A_0^A$ is set zero, and the boost gauge $\bar\alpha=0$ is chosen such that $\l \overset{B}{=} \partial/\partial_{\phi^0}$, the last expression reduces on $B$ to the usual $\sqrt q \theta = \partial_\l \sqrt q$.
The barred expansion is analogously defined as $\bar \theta_\lb = q^{ab} \nabla_a \lb_b$.

The \emph{tangential acceleration} $a_a$ is defined as
\begin{equation}
  a_a := q_a{}^b \nabla_\l \l_b.
\end{equation}
It vanishes on $B$. That can be seen by writing $a_a = q_a{}^c \l^b (\rd \bm \l)_{bc}$, and then employing Frobenius' theorem, or explicitly evaluating $a_a= e^{-\bar\alpha} q_a{}^b \nabla_b \beta \overset{B}{=} 0$.
Since $a_a \overset{B} = 0$ and also $\l^a \nabla_\l \l_a = 0$ because $\l$ has constant modulus, we obtain that $\nabla_\l \l_a$ on $B$ must be parallel to $\l_a$: $\l$ is geodesic. The proportionality factor is the \emph{normal acceleration} 
\begin{equation}
 \gamma := \lb^a \nabla_\l \l_a.
\end{equation} 
Although $\lb$ is in general not geodesic, we introduce the ``barred'' normal acceleration
\begin{equation}
\bar\gamma := \l^a \nabla_\lb \lb_a.
\end{equation}
Lastly, we introduce the \emph{twists} $\eta_a$ and $\bar\eta_a$, and the \emph{normal connection} $\omega_a$. 
\begin{align}
 \eta_a :={}& - q_a{}^b \nabla_\lb \l_b \nonumber \\
 \bar\eta_a :={}& - q_a{}^b \nabla_\l \lb_b \nonumber \\
 \omega_a :={}&  q_a{}^b \lb^c \nabla_b \l_c
\end{align}
The combination $\eta - \bar\eta$ which computes the commutator $q_{ab}[\l,\lb]^b$ is essentially the curvature of the $\text{Diff}(S)$--connection:
\begin{equation}
\eta^a - \bar\eta^a = e^{-h} [D_0, D_1]^a.
\end{equation}
This completes our geometrical setup. Let us note that under the boost transformations $(\l, \lb)\rightarrow(e^\epsilon \l, e^{-\epsilon}\lb)$, the tensors $(\eta, \bar\eta, \theta_\ell,\bar\theta_{\bar\l}) $
transform covariantly to become $(\eta, \bar\eta, e^{\epsilon}\theta_\ell,e^{-\epsilon} \bar\theta_{\bar\l})$, while the coefficients $(\gamma,\bar{\gamma},\omega_a)$ transform inhomogeneously as connections and become
 $(e^\epsilon(\gamma+\nabla_\ell\epsilon),e^{-\epsilon}(\bar{\gamma}-\nabla_{\lb}\epsilon),(\omega_a + q_a{}^b \nabla_b \epsilon))$.
We now turn to our main task of evaluating the symplectic potential on a null hypersurface.

\section{The Symplectic Potential on a Null hypersurface}
\label{sec:Theta_null}

The symplectic potential current integrated on $B$ is
\begin{equation}\label{eq:Theta_starting}
\Theta_B = 
- \int_B \Big(  e^{\bar\alpha} \Theta^a \l_a \Big) \rd B = 
 \frac12 \int_B \Big(  e^{\bar\alpha} \big(\nabla_\l \delta g- \l_a \nabla_b \delta g^{ab}  \big) \Big)\rd B,
\end{equation}
where we have used  the expression (\ref{eq:Theta_general}) for the symplectic current and our expression (\ref{eq:volume_form_B}) for the pullback of the volume $(D-1)$--form. We also introduced the abbreviation $\rd B = \sqrt q   \rd \phi^0 \wedge \rd \sigma^2 \wedge...\wedge \rd \sigma^{D-1}=  \rd \phi^0 \wedge \rd S$ for the volume form on $B$. $\rd B$ depends on  $q$ and its variation is given by 
\be
\delta \rd B =  \frac12 \delta q \, \rd B.
\ee
Let us first evaluate\begin{equation} \label{eq:Theta_l}
-\Theta^a \l_a = \frac12 \big(\nabla_\l \delta g- \l_a \nabla_b \delta g^{ab} \big),
\end{equation}
using the decomposition of variations and the extrinsic geometry introduced in section \ref{sec:setup}. 

\subsection{Evaluation of $\Theta^a \l_a$}
The second term $-\frac12 \l_a \nabla_b \delta g^{ab}$ of the last equation requires some work. We integrate it by parts, and using that $\delta g_{ab} \l^b = \delta \l_a - g_{ab} \delta \l^b$ obtain:
\begin{equation}\label{eq:theta_messy_term}
-\frac12 \l_a \nabla_b \delta g^{ab} = \frac12 \big( \delta g_{ab} \nabla^a \l^b + \nabla_a (\delta \l^a - g^{ab} \delta \l_b) \big).
\end{equation}
Let us consider the last term of the last equation, and insert the completeness relation (\ref{eq:completeness}) inside the derivative.
\begin{align}
 \frac12 \nabla_a ( \delta \l^a - g^{ab} \delta \l_b) = {} & \frac12 \nabla_a \big(q^a{}_b \delta \l^b - \l^a (\lb^b \delta \l_b + \l^b \delta \lb_b) - 2 \lb^a (\l^b \delta \l_b) \big) \nonumber \\
 = {} &\frac12 \Big(  \nabla_a (q^a{}_b \delta \l^b) - (\lb^b \delta \l_b + \l^b \delta \lb_b) (\theta_\l+\gamma) - 2 (\l^b \delta \l_b) (\bar\theta_\lb+\bar\gamma) \nonumber \\
 & \qquad  - \nabla_\l (\lb^b \delta \l_b + \l^b \delta \lb_b)  - 2 \nabla_\lb (\l^b \delta \l_b) \Big).
\end{align}
In the first line we have used that the variations stays orthogonal to the surfaces $S$, i.e. $q^{ab} \delta \l_b = 0$, for the second line, we have used $\nabla_a \l^a = \theta_\l+\gamma$ and $\nabla_a \lb^a = \bar\theta_\lb+\bar\gamma$.

The first term in (\ref{eq:theta_messy_term}) is $\delta g_{ab} \nabla^a \l^b$. It is already of the form $P \delta Q$. To evaluate it, we insert the decomposition of the metric twice. Comparing with the projected variations and the definitions of extrinsic geometry from section \ref{sec:setup}, it becomes the sum of six terms which are not identically zero:
\begin{align}
 \delta g_{ab} q^{ac} q^{bd} \nabla_c \l_d ={} & \delta q_{ab} \theta_\l^{ab} \nonumber\\
 \delta g_{ab} q^{ac} \l^b \lb^d \nabla_c \l_d ={}& -q^a{}_b \delta \l^b \omega_a \nonumber\\
 \delta g_{ab} \l^a \lb^c q^{bd} \nabla_c \l_d ={}& q^a{}_b \delta \l^b \eta_a \nonumber\\
 \delta g_{ab} \l^a \lb^c \l^b \lb^d \nabla_c \l_d ={}& - 2 \l^a \delta \l_a \bar \gamma \nonumber\\
 \delta g_{ab} \lb^a \l^c q^{bd} \nabla_c \l_d ={}& -q^a{}_b \delta \lb^b a_a \nonumber\\
 \delta g_{ab} \lb^a \l^c \l^b \lb^d \nabla_c \l_d = {}& (\lb^a \delta \l_a + \l^a \delta \lb_a) \gamma.
\end{align}
We have used that $q^{ab}\delta\ell_b=0$ and that the remaining terms are zero because $\l^a \nabla_b \l_a = 0$. Adding this up yields
\begin{equation}
 \frac12 \delta g_{ab} \nabla^a \l^b = \frac12 \big(  \delta q_{ab} \theta^{ab}_\ell +  \delta \l^a ( \eta_a-\omega_a ) -  \delta \lb^a  a_a + (\lb^a \delta \l_a + \l^a \delta \lb_a)\gamma - 2 \l^a \delta \l_a \bar\gamma \big).
\end{equation}
Now all that is left to evaluate is the term  $\frac12 \nabla_\l \delta g$ in (\ref{eq:Theta_l}). Using (\ref{eq:delta_g}), it becomes just
\begin{equation}
 \frac{1}{2} \nabla_\l \delta g = \frac{1}{2} \nabla_\l \big(\delta q + 2 (\lb^a \delta \l_a + \l^a \delta \lb_a) \big).
\end{equation}
We add everything up to obtain
\begin{align}\label{eq:Theta_l_first} \boxed{\begin{aligned}
-\Theta^a \l_a ={}&  \frac12 \Big( \delta q_{ab} \theta^{ab}_\l + \delta \l^a (\eta_a-\omega_a) - \theta_\l (\lb^a \delta \l_a + \l^a \delta \lb_a)  \\
& + \nabla_\l \big(\delta q + (\lb^a \delta \l_a + \l^a \delta \lb_a)\big)  + \nabla_a(q^a{}_b \delta \l^b)- 2 \nabla_\lb(\l^a \delta \l_a)  \\
& - { 2} (\l^a \delta \l_a) (\bar\theta_{\lb}+2\bar\gamma) - \delta \lb^a a_a \Big).\end{aligned}}
\end{align}
Remembering that $\Theta_B = -\int_B (e^{\bar\alpha} \Theta^a \l_a ) \rd B$, this is our first version of the symplectic potential on a null hypersurface. It is obtained in a straightforward way, by integrating by parts, inserting the completeness relation, and substituting the variations and extrinsic geometry we defined.
We have not assumed $\delta \beta \overset{B}{=}0$ or $\beta \overset{B}{=}0$. This general form for $\Theta^a \l_a$ was first derived in \cite{hopfmueller16:null_bdies_cov_ham_gr}, and is new in this form in the published literature. It would be the starting point if we wanted to obtain the pre--symplectic form $\Omega_B = \delta \Theta_B$, without restricting to variations that keep $B$ null\footnote{If we wanted to allow $ \beta \neq 0$, we would have to use $\epsilon_a \vert_B = -(\frac{e^{\bar \alpha}}{1+\beta\bb} \l_a + \beta e^{\alpha} \lb_a) \rd B$, because $\bm \l$ is no longer orthogonal to $B$.}, and will be useful also in other contexts\footnote{E.g., the boundary term of the variation of the gravitational Hamiltonian for translation along $\l$ contains $\iota_\l \Theta$. (\cite{donnelly16:local_subsystems})}. In this form, the result is not suited yet to read off the canonical pairs of gravity, since it still contains derivatives of variations. It is not of the form $P \delta Q$ with the configuration variable $Q$ depending only on the geometry of $B$.

\subsection{Splitting the Symplectic Potential into Bulk, Boundary and Total Variation}\label{sec:splitting}

In the following, we will restrict attention to metric variations that keep the hypersurface $B$ null, i.e., we set
\begin{equation}\label{eq:beta_equals_zero}
 \beta \overset{B}{=} 0, \qquad \delta \beta \overset{B}{=}0.
\end{equation}
We will reserve for the future the task of treating the more general case $\delta \beta \overset{B}{\neq} 0$. 
Using those conditions, the expressions $a_a$ and $\l^a \delta \l_a$ vanish on $B$ (but not the transverse derivative $\nabla_\lb (\l^a \delta \l_a)$ of the latter expression). The vanishing of $a_a$ implies that the variation $q_{ab}\delta \lb^b$ drops out of the symplectic structure. We see that since $q_{ab} \delta \lb^b$ dropped out, the symplectic potential does not contain $\delta A_1$, see (\ref{dA1}),  hinting that $A_1$ is a gauge degree of freedom. We will comment on this later. These conditions therefore lead to a simpler expression for the symplectic potential:
\begin{align}\label{eq:ThetaBonB}
\Theta_B ={}&\frac12  \int_B  e^{\bar\alpha} \Big( \delta q_{ab} \theta^{ab}_\ell + \delta \l^a (\eta_a - \omega_a) - \theta_\ell \delta h \nonumber \\
& + \nabla_\l \big(\delta q + \delta h\big)  + \nabla_a(q^a{}_b \delta \l^b) - 2 \nabla_\lb(\l^a \delta \l_a)\Big) \rd B.
\end{align}
We have written $\delta h$ for $(\lb^a \delta \l_a + \l^a \delta \lb_a)$.

This expression, while correct and expressed in terms of the extrinsic tensors, is not fully satisfactory yet for two reasons: It is not manifestly boost gauge invariant, and it still contains derivatives of variations. From \Ref{eq:Theta_starting} we see that the integrand of $\Theta_B$ is boost invariant because the combination $e^{\bar\alpha} \l_a$ is, but in the equation above the invariance is not insured term by term.
In order to achieve this it is worthwhile to notice  that the combination $e^{\bar\alpha} \l$ enters the symplectic potential in many instances. 
We therefore introduce the boost invariant combination 
\be
L^a := e^{\bar\alpha} \l^a,\qquad L^a\pa_a = D_0 +\beta D_1 \overset{B}{=} D_0,\qquad \delta L^a \overset{B}{=} q^a{}_b \delta L^b=\delta A_0^a.
\ee
We also denote its extrinsic curvature $\theta_L^{ab}$ simply by $\theta^{ab}$, which is equal to $\theta^{ab} = e^{\bar\alpha}\theta_\l^{ab}$.
Now using the identity $(\bar\eta-\omega)_a = q_a{}^b \nabla_b\bar\alpha$, we can evaluate 
\be
e^{\bar\alpha}  \nabla_a(q^a{}_b \delta \l^b) = \nabla_a \delta L^a - 
(\bar\eta_a-\omega_a) \delta L^a.
\ee 
We can also use that $e^{\bar{\alpha}}\nabla_\lb(\l^a \delta \l_a)\overset{B}{=}\nabla_\lb( e^{\bar{\alpha}}\l^a \delta \l_a)$ since 
$\l^a \delta \l_a\overset{B}{=}0$.
The symplectic potential can then be written as 
\begin{align}
\Theta_B ={}&\frac12  \int_B   \Big( \delta q_{ab} \theta^{ab} + \delta L^a (\eta_a - \bar\eta_a) - \theta \delta h  \nonumber \\
& + \nabla_L \big(\delta q + \delta h\big)  + \nabla_a \delta L^a - 2 \nabla_\lb(e^{\bar{\alpha}} \l^a \delta \l_a)\Big) \rd B.
\end{align}
In this form all the terms are now individually boost invariant. For the last term this is due to the fact that $\l^a \delta \l_a\overset{B}{=}0$. We have also discovered that the most convenient boost gauge for the symplectic structure is $\bar\alpha = 0$, since it identifies $\l = L$. Note that the induced metric \Ref{eq:induced_metric} on $B$ is determined by $(q_{ab}, L^a)$.

The last term $\nabla_\lb(e^{\bar\alpha} \l^a \delta \l_a)$ is still problematic though. Indeed even if $\l^a \delta \l_a$ vanishes on $B$, its derivative $\nabla_\lb (\l^a \delta \l_a)$ does not, since the derivative is in a direction transverse to $B$. The challenge we are facing is to find a way to eliminate this transverse derivative.
In the case where the boundary is spacelike a similar issue arises. In \cite{Freidel:2013jfa} it is shown that it is possible to eliminate the transverse derivative by including it into the variation of the densitized extrinsic curvature, which leads to the Gibbons--Hawking term. This is therefore  exactly  the strategy we are now going to follow:
We show that it is possible to absorb the transverse derivative $\nabla_\lb (e^{\bar\alpha} (\l^a \delta \l_a))$ into  a total variation.

We can evaluate that $ e^{\bar{\alpha}} (\l^a \delta \l_a) =  {-} e^{\alpha} \delta \beta $ and therefore its transverse derivative is given by $ \nabla_\lb[ e^{\bar{\alpha}}(\l^a \delta \l_a)]
\overset{B}{=} {-} D_1( \delta \beta)$, where we used $\delta \beta \overset{B}{=}0$.
Note that (even outside of $B$) the normal acceleration can be written as 
$\gamma {=} e^{-h} [D_0 e^\alpha + D_1( e^\alpha \beta)]$. This suggests that we can extract from its variation
the transverse derivative up to tangential derivatives.
Before doing so, one has to remember that the normal acceleration transforms as a connection under boosts, while we want to keep boost invariance manifest.  Under the rescaling $(\l, \lb) \rightarrow (e^{\epsilon} \l, e^{-\epsilon} \lb)$, $\gamma$ transforms as
\begin{equation}
\gamma \rightarrow e^{\epsilon} ( \gamma+ \nabla_\l \epsilon).
\end{equation}
This suggests to introduce the {\it surface gravity} which is the boost invariant combination
\be\label{eq:def_kappa}
\kappa:= e^{\bar{\alpha}}(\gamma +\nabla_\l \bar\alpha). 
\ee
It is boost invariant, since the transformation of $\bar\alpha$ and $\nabla_\l \bar\alpha$ cancels the non--invariant terms in $\gamma$.
It corresponds to the normal acceleration $\kappa= \bar L_a\nabla_L L^a$ of the vector $L{=}D_0+\beta D_1 $.
Using metric parameters, the surface gravity $\kappa$ can be written as
\be
\kappa    \overset{B}{=} D_0 h + D_1 \beta,
\ee  and is manifestly boost gauge invariant (see appendix A).

In appendix B, we calculate the total variation of the surface gravity on $B$ for variations that preserve the nullness of $B$, i.e., such that $\delta \beta\overset{B}{=}0$. It is given by 
\begin{align}\label{eq:delta_kappa}
\delta \kappa  \overset{B}{=}{}& \delta L^a (\eta_a + \bar\eta_a)
+ \nabla_L \delta h   - \nabla_\lb (e^{\bar\alpha}\l^a \delta \l_a). 
\end{align}
By using these results we can now write down the symplectic potential in a form intrinsic to $B$ which does not involve any transverse derivatives. It reads
\begin{align}
\Theta_B ={}&\frac12  \int_B   \Big( \delta q_{ab} \theta^{ab} - \delta L^a (\eta_a +3 \bar\eta_a) - \theta \delta h \nonumber \\
& + \nabla_L \big(\delta q - \delta h\big)  + \nabla_a \delta L^a + 2 \delta \kappa \Big) \rd B.
\end{align}
In order to finalize the expression we first need to integrate by parts the derivative along $L$, producing a total derivative. We use that for any $\rho$,
\be
{\sqrt q} \nabla_L \rho
\overset{B}{=}
 \pa_a \left[\sqrt q D_0^a \rho \right]
-{\sqrt q}\, \theta  \rho.
\ee
where we used that $L^a \overset{B}{=}  D_0^a$ and that 
$\partial_a (\sqrt q D_0^a) = \sqrt q \theta$.
We can also express divergences of vectors tangential to $S$ as 
\be
\sqrt{q} \nabla_a \delta L^a = \pa_a(\sqrt q \delta L^a) + \delta L^a (\eta_a +\bar\eta_a).
\ee
These identities are proven in appendix \ref{ap:bder}.
We also need to convert the last term into a total variation, using that $\delta (2\kappa \rd B) =
(2\delta \kappa + \kappa \delta q) \rd B$.
This gives us 

\begin{align}\label{eq:ThetaBonBB} 
\Theta_B ={}&\frac12  \int_B   \Big( \delta q_{ab} \theta^{ab} - 2\delta L^a \bar\eta_a - (\kappa+ \theta) \delta q \Big) \rd B \nonumber \\
& + \frac12  \int_{\pa B}  \Big( L^a \big(\delta q  
-\delta h \big)  +  \delta L^a     \Big) \rd_a S
+\delta \Big( \int_B   \kappa\,  \rd B\Big) .
\end{align} 
Here we defined the directed volume element on $\p B$ as $\rd_a S :=\iota_{a} \rd B$. In particular, in the foliation frame we have $\rd_0 S =\rd S$. 

This expression is the sum of three terms, a bulk symplectic potential, a boundary symplectic potential and a total variation.
The variational terms in the bulk symplectic potential only involve $\delta q_{ab}$ and $\delta L^a$, which form the intrinsic geometry of $B$. In particular 
we see that the term involving the variation $\delta h$ has cancelled from the bulk part.
This term is still present in the boundary contribution of the symplectic potential. In order to remove it we introduce another total variation 
\bea
- \delta h\,L^a  \rd_a S &=& 
-\delta\big( h L^a\rd_a S \big)
+
\left[\delta L^a h  +  \tfrac12 h \delta q  L^a \right] \rd_a S,
\eea
where we have used $\delta \rd_a S = \frac12 \delta q \rd_a S$.
Finally we get
\begin{align} \label{eq:Theta_final} \boxed{
\begin{aligned}
\Theta_B ={}&\frac12  \int_B   \Big( \delta q_{ab} \theta^{ab} - 2\delta L^a \bar\eta_a - (\kappa+ \theta) \delta q \Big) \rd B \\
& + \frac12  \int_{\pa B}  \Big( \big(1+\tfrac12h\big) L^a \, \delta q   +  \big(1+h \big)\delta L^a     \Big) \rd_a S\\
&+\delta \Big( \int_B   \kappa\,  \rd B - \frac12 \int_{\pa B}  h L^a\rd_a S \Big)  .
\end{aligned} }
\end{align}
This is the final expression we were looking for. 

It turns out that the boundary and the total variation part of the symplectic potential can be written in a variety of different ways: First using  $\sqrt q \theta = \p_a (\sqrt q D_0^a)$ and $D_0 \overset{B}{=}L$, it is important to note that the expansion $\theta$ is a boundary term on $B$:
\begin{equation}
\int_B \theta \rd B = \int_{\p B} L^a \rd_a S.
\end{equation}
The variation of the last equation becomes
\begin{equation} \label{eq:delta_int_theta}
\delta \Big(\int_B \theta \rd B \Big) = \delta \Big(\int_{\p B} L^a \rd_a S\Big) = \int_{\p B} (\delta L^a +\tfrac12 \delta q L^a) \rd_a S.
\end{equation}
Let us note that $\nabla_a L^a = \theta + \kappa$, and  anticipate that the total variation in $\Theta_B$ can be reabsorbed into a Lagrangian  boundary term. We see that we can express $A_B^{\text{bulk}}$ as the integral of $\nabla_a L^a$, which is the null analog of the Gibbons--Hawking term which features the divergence $K = \nabla_a N^a$ of the unit normal to the hypersurface (a boundary term of this form is given in \cite{parattu16:unified_bdy_term}). In that form, our expressions read
\begin{align}
\Theta_B ={}& \frac12  \int_B   \Big( \delta q_{ab} \theta^{ab} - 2\delta L^a \bar\eta_a - (\kappa+ \theta) \delta q \Big) \rd B + \frac12 \int_{\p B} \big( \tfrac12 h L^a \delta q + (h-1) \delta L^a \big) \rd_a S \nnn
& + \delta \Big(\int_B (\theta+\kappa) \rd B - \frac12 \int_{\p B} h L^a \rd_a S \Big).
\end{align}
Below, we write our result in a different way that we expect to be more adapted to the study of boundary degrees of freedom. Starting from (\ref{eq:Theta_final}) and using that 
\be
\delta\big( L^a\rd_a S \big) =
\left[\delta L^a   +  \tfrac12  \delta q  L^a \right] \rd_a S, 
\ee we express the symplectic potential as the sum of a bulk term, a boundary term, and the variations of a boundary action and a corner action:
\be\label{Thetafull}
\Theta_B = \Theta_B^{\text{bulk}} + \Theta_{\partial B} + \delta A_B + \delta a_{\partial B},
\ee
where 
\begin{align}\label{eq:Theta_terms}\boxed{\begin{aligned}
\Theta_B^{\text{bulk}} ={}& 
\frac12 \int_B   \big(   \theta^{ab} \delta q_{ab} - 2 \delta L^a \bar\eta_a - (\kappa+\theta) \delta q \big) \rd B \\
\Theta_{\partial B} ={}&  \frac12  \int_{\pa B}  \Big(  \tfrac12\big(1+h\big) L^a \, \delta q   +  h\delta L^a     \Big) \rd_a S   \\
 A_B ={}& \int_B \kappa\,  \rd B \\
a_{\partial B} ={}&\frac12 \int_{\pa B}  \big(1-h\big) L^a\rd_a S .\end{aligned} }
\end{align}
We analyze these expressions in the next sections.
%
\section{Canonical Pairs}\label{sec:canonical_pairs}

We now read off the null canonical pairs of gravity from \Ref{eq:Theta_final}, comparing with the schematic expression \Ref{eq:Theta_schematic}.
The momenta for the trace $\delta q=\delta q_{ab} q^{ab}$ and the traceless part $\delta q_{\langle ab\rangle}$ of the variation of the induced metric on $S$ have different forms. It is therefore natural to split the induced metric into a conformal part and the volume element. We define the \emph{conformal induced metric} on $S$, which has unit determinant:
\begin{equation}
 \tilde q_{ab} := |q|^{-\frac{1}{D-2}} q_{ab}.
\end{equation}
Its variation $\delta \tilde q_{ab} = |q|^{-\frac{1}{D-2}} (\delta q_{ab} - \frac{1}{D-2} \delta q\ q_{ab})$ is traceless. Its momentum is the \emph{conformal shear}
\begin{equation}
\tilde\sigma^{ab} = |q|^{\frac{1}{D-2}} (\theta^{ab} - \tfrac{1}{D-2} q^{ab} \theta) = - \frac12 q^a{}_{a'} q^b{}_{b'} \pounds_L \tilde q^{a'b'},
\end{equation}
which is also traceless, and captures the change of the conformal inverse metric $\tilde q^{ab} = |q|^{\frac{1}{D-2}} q^{ab}$ along $L$.
Splitting the term $\delta q_{ab} \theta^{ab}$ into its trace and traceless parts then yields
\begin{equation}
\delta q_{ab} \theta^{ab} = \delta \tilde q_{ab} \tilde \sigma^{ab} + \tfrac{1}{D-2} \delta q\ \theta.
\end{equation}
These replacements give
\begin{align}
\Theta_B^{\text{bulk}} ={}& \int_B \left( \frac12 \delta \tilde q_{ab} \tilde \sigma^{ab} - \left(\tfrac{D-3}{D-2} \theta +\kappa\right) \delta \ln \sqrt q - \delta L^a \bar\eta_a\right) \rd B,
\end{align}
where we have substituted $\delta q = 2 \delta \ln \sqrt q$ to produce an exact variation.
From this we read off the bulk canonical pairs in the following table:
\be\label{tab:null_canonical_pairs}
\begin{array}{|lc|cl |}
\hline 
\text{Bulk  configuration}&  \hspace{1cm} & \text{Bulk  momentum} & \\ 
\hline & & & \\
\text{Conformal metric:} & \tilde q_{ab}  \,\,\,\,&  \frac12 \tilde \sigma^{ab}  &\text{Conformal shear}\\  & & &\\
\text{Normal  vector:} &  L^a  \,\,\,\,& - \bar\eta_a & \text{Twist}\\
& & & \\ 
\text{Volume  element:} &\ln \sqrt q \,\,\,\,&  - \big(  \kappa+ \tfrac{D-3}{D-2} \theta  \big) &\text{Expansion, surface gravity} \\
& & &\\
\hline
\end{array}
\ee
Note that what we call momenta are $B$-densities $P\rd B$.
The boundary canonical pairs can be read off from \Ref{eq:Theta_terms} and are
\begin{align}
\left( L^a, \tfrac12 h q_a{}^b\rd_b S \right) \text{ and }\left(\ln \sqrt q, \tfrac12 (1+h) L^a \rd_a S \right).
\end{align}
We postpone their detailed analysis to a future publication. 

We recover the surprising fact that the configuration variables in the bulk of $B$ contain only variations of the induced metric \Ref{eq:induced_metric} on $B$ and no variations of the normal metric. That is analogous to the spacelike and timelike symplectic structure, and was not obvious from the outset. In the null case, also the configuration variables on the corner $\p B$ are a subset of the induced geometry.
It is well known that the momentum conjugate to the conformal metric is the shear. More interesting and surprising are the spin--1 and spin--0 momenta $-\bar\eta_a$ and $-(\kappa + \tfrac{D-3}{D-2} \theta)$. Let us analyze them.

The spin $1$ configuration variable is $L$. Since its $\p_{\phi^0}$--component is fixed, its variation is $\delta L^a = \delta A_0^a$ and is purely tangential to $S$. The momentum $\bar{\eta}$ conjugate to $L$ is given by $\bar\eta_a = -q_a{}^b \nabla_\l \lb_b$. Since $\lb$ determines the orientation of $S$ within $B$, $\bar\eta_a$ describes how the cross--sections $S$ of $B$ tilt and twist when parallel--transporting them along $\l$.  It can be expressed as the sum of two terms (see appendix \ref{ap:extrinsic}):
\begin{equation}
\bar\eta_A = \frac12 \big( \partial_A h - F_A\big), \qquad F_A := q_{AB} e^{-h} \left(\pa_0 A_1^B- \pa_1 A_0^B+ [A_0,A_1]^B\right)
\end{equation}
Here $F_A$ measures the non-integrability of the normal two planes. Indeed it is equal to $q \cdot [L,\bar{L}]$, which vanishes only if the normal two planes are integrable.
The other term measures the rate of change of the redshift factor $h$ along the cross--section $S$.

Using the boost gauge $\bar\alpha=0$, Damour (\cite{damour82:surface}) first interpreted $\omega_a = q_a{}^b \lb^c \nabla_b \l_c$ as a momentum density. He was motivated by the fact that for a cylindrically symmetric black hole, an integral of $\omega_a$ is the total angular momentum, and that in the Navier--Stokes--like equation $q_a{}^b L^c R_{bc} = 0$, $\omega_a$ plays the role of a linear momentum. However, $\omega_a$ is not boost gauge invariant and transforms as a connection under the boost gauge. The twist $\bar \eta$ is boost gauge invariant and coincides with $\omega$ in the boost gauge $\bar\alpha = 0$ since $\bar\eta_a - \omega_a = q_a{}^b \nabla_b \bar\alpha$. The twist $\bar\eta$ is thus the proper boost gauge invariant generalization of $\omega$.
In the light of the fluid interpretation of null surfaces, it is thus very natural that we found $\bar\eta$ as the momentum conjugate to the ``displacement'' $A_0$. We have confirmed Damour's interpretation of $\omega$ from a symplectic analysis.

The spin--0 momentum $-(\kappa + \tfrac{D-3}{D-2} \theta)$ conjugate to $\ln \sqrt q$ is a dimension dependent combination of the expansion $\theta$ and the surface gravity $\kappa$. The surface gravity $\kappa$, which can be defined through $\nabla_{L} L_a =\kappa L_a$, is given as the sum of two terms:
\be 
\kappa = D_0 h + D_1 \beta. 
\ee
We will see in more detail that this acceleration is the sum of an inertial acceleration term $D_0 h$ with $h$ playing the role of a velocity, and a Newtonian acceleration term $D_1 \beta$ with $\beta$ playing the role of the Newtonian potential. In the case of a non--expanding null surface ($\theta=0$) we recover the pair from black hole thermodynamics: the volume element $\sqrt q$ is conjugate to the surface gravity $\kappa$.

In appendix \ref{ap:extrinsic}, expressions for the momenta are derived in terms of the metric parameters $(A_i^A, q_{AB}, h, \beta, \bb)$. 
The bulk momentum conjugated to the conformal metric is 
\be
\tilde{\sigma}^{AB} = \tilde{\theta}^{\langle A B\rangle},\quad {\mathrm{with}} \quad \tilde{\theta}^{ A B}=
-\frac12\pa_0\tilde{q}^{AB} +  \frac12 \big( \tilde{q}^{ A C} \partial_C A_0^{ B } + \tilde q^{CB} \partial_C A_0^A -  A_0^C\pa_C\tilde{q}^{AB})
\ee
where $\langle A B\rangle$ denotes the traceless part. Note that if one introduces $\delta_A$ to be the two dimensional covariant derivative compatible with $q_{AB}$ and denotes $\tilde\delta^A =\tilde{q}^{AB} \delta_B$ this can be also simply expressed as 
\be
  \tilde{\theta}^{ A B}=
-\frac12\pa_0\tilde{q}^{AB} +  \frac12 \big( \tilde{\delta}^A A_0^{ B } + \tilde{\delta}^B A_0^A).
\ee
The first contribution comes from the time dependence of $q_{AB}$ while the second contribution is analogous to the rate of strain $\pa^{(A} v^{B)}$ familiar in hydrodynamics if we identify  $A_0^A$  as the `velocity' $\dot{\sigma}^A$ of non-rotating observers on $S$ which follow the integral curves of $\l$.

From the expressions for the momenta in terms of metric parameters, two transformations are apparent that are gauge in the sense that they are in the kernel of the symplectic form. 
Firstly, once we fix $\beta \overset{B}{=}0$, the parameter $A_1$ enters the symplectic potential only through the curvature $F_A$ (see also discussion after \Ref{eq:beta_equals_zero}). More precisely, the only term in $\Theta_B$ that contains $A_1$ is the term $F_A \delta A_0^A$. It can be easily seen that this term is invariant under the gauge transformations that we introduced in \Ref{eq:normal_connection}, 
\be 
\delta_V A_0^A = \p_0 V^A + [A_0, V]^A ,\qquad \delta_V A_1^A = \p_1 V^A + [A_1, V]^A.
\ee
Since the transformation $\delta_V$ does not affect any other term in the symplectic potential, it is pure gauge. The transformation can be used to control $A_1^A$, without affecting the rest of the analysis.
Similarly, we see that the metric parameter $\bb$, which was introduced in \Ref{eq:l_form} and depends on the choice of the observers $\phi^0$, does not appear in the symplectic potential at all, neither in the configuration variation nor in the momentum variables. This is also a consequence of the fact that we have restricted to variations that keep $B$ null, i.e., obey $\delta \beta = 0$. We can therefore choose $\bar\beta$ arbitrarily without affecting the analysis (as long as $\bar{\beta}>0$), a fact we will exploit in the next section.

\subsection{Normal frames, redshift factor and surface gravity}\label{redshift}
As we have just seen the parameter $\bar{\beta}$ is at our disposal. Changing $\bar{\beta}$ can be achieved by a change of foliation that affects only the time foliation $\phi^0$.  This can be shown explicitly by considering a diffeomorphism parallel to $D_0$, associated with the foliation transformation 
\be
\delta_Y \phi^1=0,\qquad \delta_Y \phi^0= e^{-h} Y,\qquad 
\delta_Y \sigma^A = e^{-h} Y A_0^A.
\ee
We can check that $h$, $\beta$, and $A_1$ are unchanged on $B$ if $Y$ is chosen to vanish on $B$, which we now do.  The only change of the normal metric then comes from a transformation of $\bar\beta$, given by $\delta_Y\bar\beta \overset{B}=e^{-h} D_1 Y$. And we see that fixing $\bar\beta$ can be achieved by chosing the $\phi_0$ foliation appropriately.
There exist several special values for the parameter $\bar \beta$ that are of physical interest, because they give back generalizations of various coordinate systems for the Schwarzschild metric: Schwarzschild coordinates, Eddington--Finkelstein coordinates or Painlev\'e--Gullstrand coordinates.

Schwarzschild type coordinates are obtained by chosing $\beta^{-1}=\bar{\beta}$. In this case the normal metric is diagonal, and reads
\be
H_{ij}\rd\phi^i\rd\phi^j= e^h[-\beta (\rd \phi^0)^2 + \beta^{-1} (\rd \phi^1)^2].
\ee
We can also choose Eddington--Finkelstein type coordinates\footnote{In order to get the Schwarzschild black hole metric in Eddington--Finkelstein coordinates we need to choose $v = \phi^0+ a\phi^1$ as an ingoing null coordinate, $r=\phi^1$ as a radial coordinate, and set $h = 0$ and $\beta$ such that $(2 \beta) / (1 + a \beta) = 1 - 2GM/r$.} which are characterized by choosing $\bar{\beta}=a$, where $a$ is a constant. In this case the normal metric reads
\bea
H_{ij}\rd\phi^i\rd\phi^j
&=&e^{h}\left[ - \frac{2\beta }{1+ a\beta}\rd(\phi^0+a\phi^1)^2 + \rd\phi^1 \otimes \rd (\phi^0+a\phi^1) \right].
\eea
In analogy to Eddington--Finkelstein coordinates, the vector $D_1$ is null for that metric, and the surfaces $\phi^0 + a \phi^1 = \const$ are null.

Finally, Painlev\'e--Gullstrand type coordinates are obtained when one chooses $\bar\beta$ such that $g^{-1} (\rd \phi^0, \rd \phi^0) = -1$, i.e., 
\be\label{PG}
\bar{\beta} = \frac{e^h}{( 2 -\beta e^h)}
.
\ee
%
For this choice the normal metric\footnote{To recover Schwarzschild with $\phi^1=r$ we need to impose $h=0$ and $\beta = 1-\sqrt{2GM/r}$.} corresponds to 
\begin{align}
H_{ij}\rd\phi^i\rd\phi^j={}&- \left(\rd \phi^0\right)^2 
+ e^{2h}\left( \rd \phi^1-(\beta -e^{-h} ) \rd \phi^0\right)^2.
\end{align}
It is easy to see that as for Painlev\'e--Gullstrand coordinates, the family of observers $v = - g^{-1}(\rd \phi^0)$ follow affinely parametrized geodesics, with proper time $\phi^0$. Let us analyze this frame in more detail.

The velocity of  the free--falling observers $v$ is explicitly given by
\be
 v= D_0 + (\beta -e^{-h}) D_1. 
\ee
In other words, restricting to the normal plane of $S$ the velocity of free falling observers is given by $v_{\mathrm FF} = \dot\phi^1= \beta -e^{-h}$.
We see that all the observers that have $\beta< e^{-h}$ are radially moving inwards. In this frame, light orthogonal to $S$ travels along out-going or in-going curves respectively given by 
 \be\label{eq:PG_speedoflight}
 v_+ = \dot\phi^1= \beta ,\qquad v_- = \dot\phi^1= (\beta -2 e^{-h}),
 \ee
where $v_+$ denotes the  outgoing velocity and $v_-$ the ingoing light rays respectively. 
This means that the speed of light as measured by the freely falling observers $v$ in this geodesic frame is given by $\pm e^{-h}$. In particular we have that 
\be\label{light}
v_\pm =\frac12\left( v_{FF} \pm c \right),\qquad c = e^{-h}. 
\ee We can thus interpret the speed of light $c=e^{-h}$ as encoding the redshift. More precisely, let us asssume that we fix the  remaining freedom
(which is physical by our analysis) in the foliation $\phi^1$ to set $\beta = 0$ everywhere. Then, the surfaces $\phi^1 = \const$ are null, and $\phi^0$ is the proper time of the geodesic observers $v = -g^{-1}(\rd \phi^0)$. This frame is called the geodesic lightcone frame, and is used in cosmology in order to define cosmological averaging \cite{gasperini11:lightcone_averaging, Fanizza:2013doa}. The normal metric then reads
\begin{equation}
H_{ij} \rd \phi^i \rd \phi^j = e^h \rd \phi^0 \otimes \rd \phi^1 + e^{2h} (\rd \phi^1)^2.
\end{equation}
In this frame, the static observers are moving at the speed of out-going light.
In this frame, the frequency of light the observer measures is given by the (negative) scalar product between the affinely parametrized null generator $\l \overset{B}=g^{-1}(\rd \phi^1) $ and the velocity $v$ of geodesic observers, up to a normalization that is constant on each light ray:
\be
 \nu= - \l^a v_a = e^{-h}.
\ee
The redshift between source and observer is then given by
 \be
 (1+z):=\frac{\nu_s}{\nu_o} = e^{h_0 - h_s}.
 \ee 
We see that $e^{-h}$ indeed encodes the redshift. That justifies the term ``redshift factor'' for $h$.

\begin{figure}[h]
{\centering
\includegraphics[width=.6\textwidth]{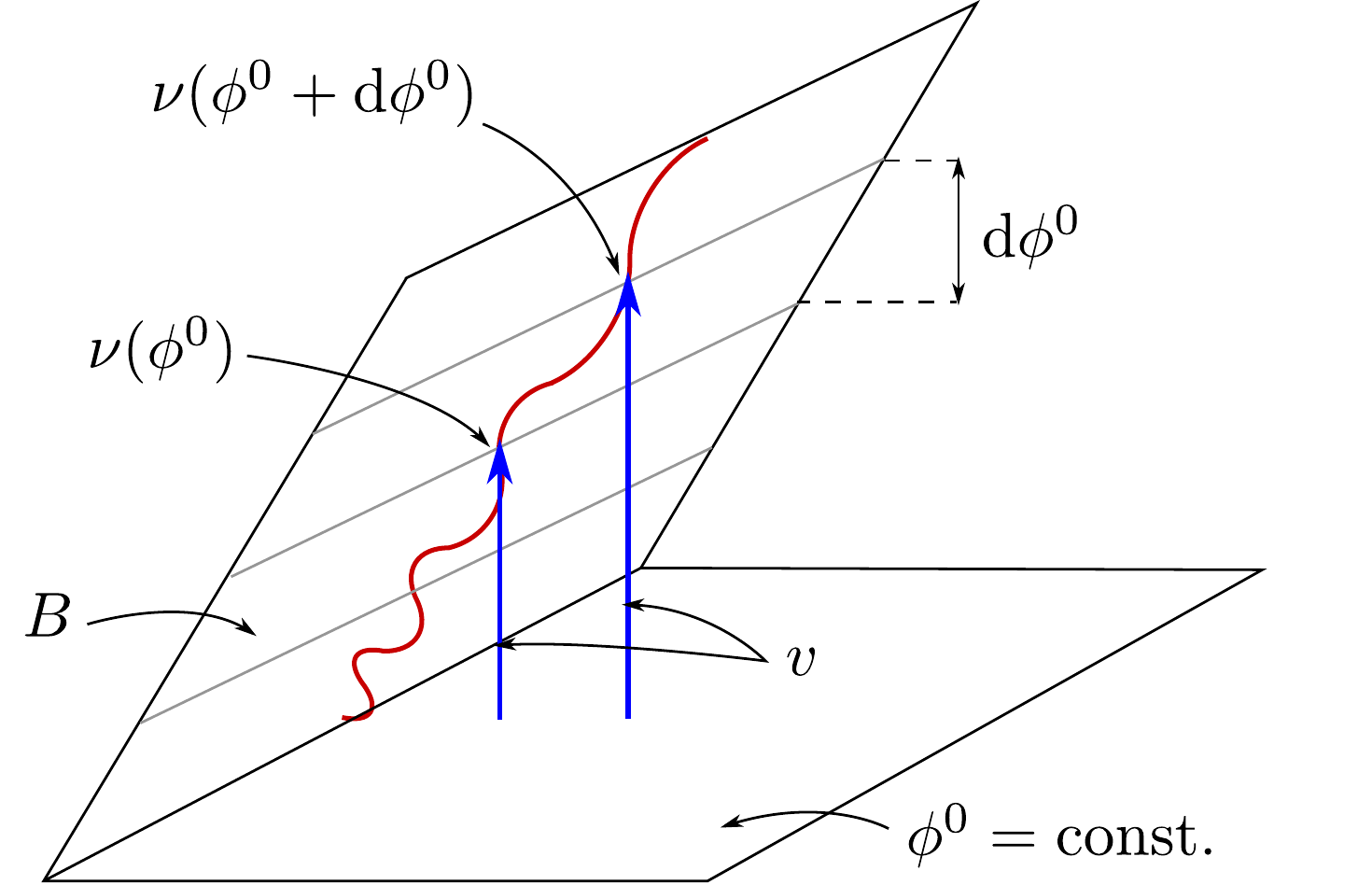}

}
\caption{The observer $v = - g^{-1}(\rd \phi^0)$ crosses the null surface $B$, and measures the frequency $\nu$ of light rays propagating along $B$. The redshift, i.e., the relative change of frequency, per unit time $\phi^0$ is given in the geodesic lightcone frame by: $\displaystyle
z = \big(\nu(\phi^0) - \nu(\phi^0  + \rd \phi^0)\big) / \nu(\phi^0) = D_0 h \cdot \rd \phi^0$
}
\label{fig:redshift}
\end{figure}

The surface gravity $\kappa$, which appears in the spin--0 momentum and the boundary action, is the sum of two terms:
\be 
\kappa = D_0 h + D_1 \beta. 
\ee
By a Newtonian analogy we can see this as the sum of an inertial acceleration term $D_0 h$ and a Newtonian acceleration term $D_1 \beta$, which we will interpret in different frames. To interpret the term $D_0 h$, we again use the geodesic lightcone frame. $D_0 h$ is the relative change of frequency per unit time, or the infinitesimal redshift, along a light ray (see figure \ref{fig:redshift}). Loosely speaking, it is thus the acceleration of a lightlike observer following lightrays perpendicular to $S$, relative to the geodesic observer $v$. We can then interpret it as the inertial acceleration term for that observer.

To interpret the term $D_1 \beta$, let us start from Painlev\'e--Gullstrand type coordinates, and use the freedom in the foliation field $\phi^1$ to set $h=0$. Then, from \Ref{light} we see that the speed of light is $1$ in that frame. We call this frame the Galilean light frame. On $B$, the acceleration of the geodesic observer $v$ relative to the static observer in the frame $(\phi^0, \phi^1)$ is then given by
\begin{equation}
v^a \p_a v^1 = - D_1 \beta.
\end{equation}
This is the expression for a radial acceleration with the radial coordinate $\phi^1$, with $\beta$ taking the role of the Newtonian potential. It is consistent with the standard Newtonian limit of general relativity because $\beta = -\frac12 g_{00}$ in the Galilean light frame.

\section{Lagrangian Boundary Terms}\label{sec:lagrangian_bdy}

In the expression (\ref{Thetafull}) for the symplectic potential, we have extracted a total variation from the symplectic potential. That corresponds to a choice of polarization: It tells us which are the configuration and which the momentum variables. This can be seen most clearly by noting that such a total variation can be used to interchange to roles of configuration and momentum variables:
\begin{equation}
\Theta = P \delta Q = - Q \delta P + \delta (PQ).
\end{equation}
The choice of polarization we have made is that the configuration variables $Q$ should not contain derivatives of the metric. 
From \Ref{eq:def_eq_Theta} one sees that the addition of a boundary term in the Lagrangian will modify the symplectic potential $\Theta$ by a total variation (up to the space--time closed ambiguity):
\begin{align}
L \rightarrow L + \rd A \Rightarrow 
\Theta \rightarrow \Theta + \delta A.
\end{align}
The total variation in $\Theta_B = \Theta_B^{\text{bulk}} + \Theta_{\p B} + \delta (A_B + a_{\p B})$ can thus be cancelled by adding a boundary term and a corner term to the action: 
\be
 S = \int_M L - A_B - a_{\p B}
\ee
From \Ref{eq:Theta_terms}, we therefore make the following suggestion for the action of a space--time region with null boundaries which may possess corners:
\be \boxed{
 S = \frac12 \int_M R\epsilon - \int_B \kappa \rd B - \frac12 \int_{\p B} \big(1 - h) L^a \rd_a S.
} \ee
Note that the corner term vanishes for segments of $\p B$ that contain the null direction $L$, since in that case $L^a \rd_a S = 0$.

A similar line of reasoning to this section was followed in the recent paper by Lehner et al. \cite{Lehner:2016vdi}, and a proposal for the boundary and corner action for null boundaries was given. We have thus reproduced one of the result there with a different calculation.\footnote{Up to the summand $1$ in the corner term, which is a choice of corner polarization, see discussion after \Ref{eq:Theta_final}.} There, the null surface $B$ is taken to have cylindrical topology, with past and future boundaries that are members of the family of surfaces $S$. The boost gauge $\bar\alpha=0$ is chosen, and the coordinates $\sigma$ are chosen to be constant along the null generators of $B$, fixing $A_0 = 0$ and $\delta A_0 = 0$ from the outset. Our calculations confirm that $\bar\alpha=0$ is the most convenient gauge, because in that gauge the boundary action $\kappa$ is indeed the inaffinity of the null generators. Furthermore, we have disentangled the normalizations of the configuration variable $L$ and the null generators $\l$, and allowed for general topologies of $B$.

Parattu et al.\ (\cite{parattu16:bdy_term_grav_action_null}) also gave a suggestion for the boundary action and the canonical structure. They mostly work in the boost gauge $\alpha = 0$, and extract a total variation containing the normal acceleration $\gamma$ rather than the surface gravity $\kappa$. As can be seen from \Ref{eq:def_kappa}, $\gamma$ and $\kappa$ are inequivalent unless $\bar\alpha=0$. For that reason they obtain an extra canonical pair on $B$, which contains a piece of normal geometry as a configuration variable. As we have seen, that pair can be removed by choosing $\kappa$ rather than $\gamma$ in the total variation extracted.

\section{Conclusion}\label{sec:conclusion}

We have set up a robust and intuitive framework for dealing with the geometry of null hypersurfaces and variational quantities on them. We used it to find the null canonical pairs of gravity from the symplectic potential current $\Theta$, without introducing any gauge fixing, and gave definitive answers especially for the spin--1 and spin--0 degrees of freedom. Included in our analysis are degrees of freedom on the boundary of the null hypersurface, about which we will have more to say in a future publication \cite{hopfmueller:soft_gravitons}. Our calculations also yielded a boundary action that includes corner terms.

Two areas come to mind where the technology and results we found can be applied. This includes, firstly, understanding the ``soft graviton modes'' of the BMS group at asymptotically flat null infinity (see, e.g., \cite{Hawking:2016msc}), which we expect to be related to the spin--0 and 1 degrees of freedom and boundary degrees of freedom. Secondly, the symplectic potential controls the flow of information. We will use this intuition to define a symplectic notion of informational horizon. That will involve going on--shell, and comparing to other notions of horizons, such as isolated horizons \cite{ashtekar00:isolated_horizons}, and may provide insight on informational quantities such as the Bousso bound \cite{Bousso:1999xy}.

\section*{Acknowledgements}
The authors thank Krishna Parattu, Luis Lehner and Rafael Sorkin for interesting discussions, and Patrick Duchstein and Pratik Rath for spotting typos.
This research was supported in part by Perimeter Institute for Theoretical Physics. Research at Perimeter Institute is supported by the Government of Canada through the Department of Innovation, Science and Economic Development Canada and by the Province of Ontario through the Ministry of Research, Innovation and Science. The authors acknowledge an NSERC discovery grant.

\appendix

\section{Extrinsic Geometry Expressed in Metric Parameters}\label{ap:extrinsic}

We relate the extrinsic geometry to the derivatives of metric parameters. The versions of these identities which hold true on $B$ where used in \ref{sec:Theta_null}. We start with the normal acceleration which is the most involved expression:
\bea
\gamma &=& \lb^a \nabla_\l \l_b = \lb^a (\pounds_\l \bm\l)_a = \nabla_\l \alpha + e^{\alpha-\bar\alpha} \nabla_\lb \beta + \frac{\bb}{1+\beta\bb} \nabla_\l \beta \nonumber\\ 
&=&e^{-\bar\alpha}[ (D_0 + \beta D_1)\alpha + D_1\beta] 
= e^{-\bar\alpha-\alpha} [D_0 e^\alpha + D_1(e^\alpha \beta)]
\overset{B}{=} \nabla_\l \alpha + e^{\alpha-\bar\alpha} \nabla_\lb \beta \label{eq:sigma_in_coords}.
\eea
From this we can evaluate the  surface gravity
\be
\kappa := e^{\bar\alpha} (\gamma + \nabla_\l\bar\alpha)  = (D_0 + \beta D_1) h + D_1 \beta.
\ee
The tangential acceleration $a$, the twists $(\eta, \bar\eta)$ and normal connection $\omega$ are given by:
\begin{align}
 a_a ={}& q_a{}^b \nabla_\l \l_b = q_a{}^b (\pounds_\l \bm \l)_b = e^{-\bar\alpha} q_a{}^b \nabla_b \beta \overset{B}{=} 0 \nonumber\\
 \eta_a + \omega_a ={}& - q_a{}^b \nabla_\lb \l_b + q_a{}^b \lb^c \nabla_b \l_c = -q_a{}^b (\pounds_\lb \bm \l)_c = q_a{}^b (\nabla_b \alpha + \frac{\bb}{1+\beta\bb} \nabla_b \beta) \overset{B}{=}  q_a{}^b \nabla_b \alpha \nonumber\\ 
 \bar \eta_a - \omega_a ={}&  - q_a{}^b \nabla_\l \lb_b + q_a{}^b \l^c \nabla_\l \lb_c = -q_a{}^b (\pounds_\l \bm \lb)_b = q_a{}^b (\nabla_b\bar\alpha - \frac{\bb}{1+\beta\bb} \nabla_a \beta) \overset{B}{=} q_a{}^b \nabla_b \bar\alpha \nonumber\\
 \eta_a - \bar\eta_a ={}& - q_a{}^b \nabla_\lb \l_b + q_a{}^b \nabla_\l \lb_b = q_{ab} [\l, \lb]^b = e^{-h} q_{ab} [D_0, D_1]^a.
\end{align}
These identities are proven by inserting the parametrizations (\ref{eq:l_form}) and (\ref{eq:null_vectors}), and executing the Lie derivatives. Linear combinations of the last three identities yield
\bea
 \eta_a + \bar\eta_a &=& q_a{}^b \nabla_b h, \nnn
\bar\eta_a &=& \frac12 \big(q_a{}^b \nabla_b h - q_{ab}e^{h} [D_0, D_1]^b\big). 
\eea

In section \ref{sec:Theta_null}, we used the identity $\partial_a (\sqrt q D_0^a) = \sqrt q \theta$. Let us prove it.
First evaluate 
\be\label{dL1}
\nabla_a L^a = (q^{ab} + \lb^a \ell^b + \l^a \lb^b) \nabla_a L_b
= \theta_L + \lb^a\nabla_\l L_a =
\theta + \kappa.
\ee
Let us evaluate the same object using now the relationship between covariant and regular derivative and that $L^a = D_0+\beta D_1$:
\bea
\nabla_a L^a &=& \frac{1}{\g} \pa_a (\g L^a)= \frac{e^{-h}}{\sqrt q } \pa_a (\sqrt q e^{\alpha+\bar\alpha} L^a)\nonumber \\
&=&\frac{1}{\sqrt q } \pa_a (\sqrt q  L^a)  + \nabla_Lh
\nonumber \\
&=&\frac{1}{\sqrt q } \pa_a (\sqrt q  (D_0^a+\beta D_1^a))  + \kappa - D_1 \beta \nonumber \\
&=& \frac{1}{\sqrt q } \pa_a (\sqrt q  D_0^a) + \frac{\beta}{\sqrt q }\pa_a(\sqrt q  D_1^a)  + \kappa . \label{dL2}
\eea
Comparing (\ref{dL2}) and (\ref{dL1}) gives  what we wanted to show:
\be 
\theta = \frac{1}{\sqrt q } \pa_a (\sqrt q  D_0^a) + \frac{\beta}{\sqrt q }\pa_a(\sqrt q  D_1^a)\overset{B}{=}
  \frac{1}{\sqrt q } \pa_a (\sqrt q  D_0^a).
\ee
Lastly, the bulk momentum for the conformal metric $\tilde q_{ab}$ is the conformal shear, which is the traceless part of the expansion of the conformal metric:
\be
\tilde \sigma^{AB} = \tilde{\theta}^{ <A B>},\text{ where } \tilde \theta^{AB} =
-\tfrac12\pa_0\tilde{q}^{AB} +  \tfrac12 \big( \tilde{q}^{ A C} \partial_C A_0^{ B } + \tilde q^{CB} \partial_C A_0^A -  A_0^C\pa_C\tilde{q}^{AB}).
\ee
The shear can also be written as $\tilde \sigma^{AB} = |q|^{1/(D-2)} \theta^{<AB>}$, with $\theta^{AB} = -\frac12\pa_0{q}^{AB} +  \frac12 \big( {q}^{ A C} \partial_C A_0^{ B } +  q^{CB} \partial_C A_0^A -  A_0^C\pa_C{q}^{AB})$.

\section{Calculation of the Variation of the Surface Gravity}\label{ap:total_variation}
Let us evaluate the total variation $\delta \kappa$ that was used in \Ref{eq:delta_kappa}. Using the coordinate expression for $\kappa$ given in the previous appendix and assuming $\delta \beta \overset{B}{=}0$, we obtain
\begin{align}
\delta \kappa \overset{B}{=} \delta \big(D_0 h + D_1 \beta\big).
\end{align}

We distribute the variation, and use that $\delta\beta \overset{B}{=}0$, that the variations $\delta D_i^a = \delta A_i^a$ are purely tangential to $S$, that $q_a{}^b \nabla_b h = \eta_a + \bar\eta_a$ and that $q_a{}^b \nabla_b \beta = 0$:
\begin{align}
\delta \kappa \overset{B}{=} D_0\delta h + \delta A_0^a (\eta_a + \bar\eta_a) + D_1 \delta \beta.
\end{align}
Substituting $\delta L^a \overset{B}{=} \delta A_0^a$, $\l_a \delta \l^a = e^{\alpha-\bar\alpha} \delta \beta$ and the coordinate expressions for $L$ and $\lb$ yields
\begin{align}
\delta \kappa \overset{B}{=} \nabla_L \delta h+ \delta L^a (\eta_a + \bar\eta_a) + \nabla_\lb(e^{\bar\alpha}\l_a \delta \l^a).
\end{align}
That is the expression we used.
\section{Calculation of Integration by Parts}\label{ap:bder}
We prove identities that we used in section \ref{sec:splitting} to integrate by parts in $\Theta_B$, producing boundary terms on $\partial B$.
We first use that for any vector $V$
\bea
\sqrt q\nabla_a V^a &=& \g e^{-h}\nabla_a V^a 
= e^{-h}\pa_a(\g V^a)
=  e^{-h}\pa_a(\sqrt q e^{h} V^a)\nonumber \\
&=&\pa_a(\sqrt q  V^a) + \sqrt q V^a \pa_a h. 
\eea
If $V^a = q^a{}_b V^b$ is a tangential vector to $S$ this means that 
\be
 \sqrt q\nabla_a V^a=\pa_a(\sqrt q  V^a) + \sqrt q V^a (\eta_a +\bar\eta_a).
\ee
If on the other hand we take $V^a =\rho L^a$ we obtain the identity
\bea
{\sqrt q} \nabla_L \rho &=&{\sqrt q} L^a \nabla_a \rho 
\overset{B}{=}{\sqrt q} D_0^a \nabla_a \rho = {\sqrt q}  \nabla_a (D_0^a\rho ) -
 \rho {\sqrt q}   \nabla_a D_0^a\big] \nonumber\\  
&=&  \pa_a (\sqrt q D_0^a\rho )- \rho \pa_a (\sqrt q D_0^a )\nonumber \\
&=& \pa_a (\sqrt q D_0^a\rho )- \sqrt q \theta.
\eea
where we used that $L^a \overset{B}{=}  D_0^a$ and that 
$\partial_a (\sqrt q D_0^a) = \sqrt q \theta$.

\bibliography{references}
\end{document}